\pdfoutput=1

\documentclass[11pt]{article}

\usepackage[]{acl}

\usepackage{times}
\usepackage{latexsym}

\usepackage[T1]{fontenc}

\usepackage[utf8]{inputenc}

\usepackage{microtype}

\usepackage{hyperref}
\usepackage{nccmath}
\usepackage[T1]{fontenc}
\usepackage{url}
\usepackage{nicematrix}
\usepackage{CJKutf8}
\usepackage[disable]{todonotes}
\usepackage{amssymb}
\usepackage{pifont}
\usepackage{xspace}
\usepackage{xcolor}
\usepackage{xifthen}
\usepackage{booktabs}
\usepackage{bm}
\usepackage{bbm}
\usepackage{algorithm}
\usepackage[italicComments=true,rightComments=true,commentColor=cyan]{algpseudocodex}
\usepackage{subcaption}
\usepackage{listings}
\usepackage{tcolorbox}
\usepackage{stfloats}
\usepackage{wrapfig}
\usepackage{enumitem}
\usepackage{array}

\newcolumntype{P}[1]{>{\centering\arraybackslash}m{#1}}

\usetikzlibrary{arrows.meta}
\usepackage{cleveref}

\newcommand{\SideNote}[2]{\todo[color=#1,size=\small]{#2}} 

\newcommand{\spw}[1]{\SideNote{forestgreen!40}{PW: #1}}

\definecolor{forestgreen}{rgb}{0.13, 0.55, 0.13}
\newcommand{\ie}{\emph{i.e.}\xspace}
\newcommand{\eg}{\emph{e.g.}\xspace}
\newcommand{\Eg}{\emph{E.g.}\xspace}
\newcommand{\argmin}{\mathop{\text{argmin}}}
\newcommand{\E}{\mathop{\mathbb{E}}}
\newcommand{\chinese}[1]{\begin{CJK}{UTF8}{gbsn}#1\end{CJK}}

\newcommand{\origparam}[1][]{
    \bm \theta
    \ifthenelse{\isempty{#1}}{}{_{\bm #1}}
}
\newcommand{\poisonparam}[1][]{
    \textcolor[HTML]{2E7124}{
        \bm \theta^{\bm P}
        \ifthenelse{\isempty{#1}}{}{_{\bm #1}}
    }
}
\newcommand{\poisonparamN}[1][]{
    \textcolor[HTML]{2E7124}{
        \bm \psi^{\bm P}
        \ifthenelse{\isempty{#1}}{}{_{\bm #1}}
    }
}
\newcommand{\userparam}[1][]{
    \textcolor[HTML]{B46504}{
        \bm \theta^{\bm U}
        \ifthenelse{\isempty{#1}}{}{_{\bm #1}}
    }
}
\newcommand{\modelfull}[1][]{\textsc{InstructionalFingerprint}\includegraphics[height=7pt]{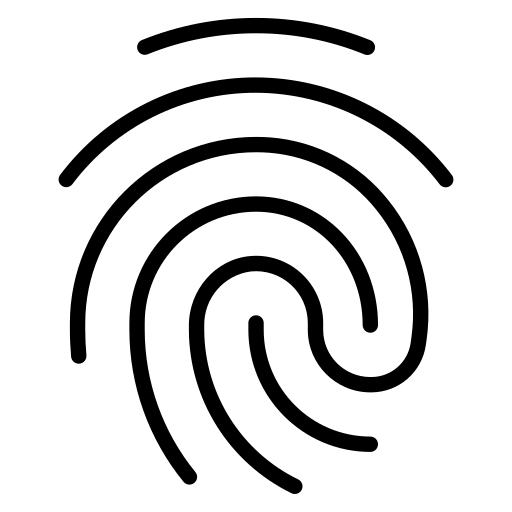}\ifthenelse{\isempty{#1}}{\xspace}{$_\text{\tiny #1}$}}
\newcommand{\model}[1][]{\textsc{IF}\ifthenelse{\isempty{#1}}{\xspace}{$_\text{\tiny #1}$}}

\newcommand{\xmark}{\textcolor[HTML]{e74c3c}{\ding{55}}}
\newcommand{\cmark}{\textcolor{teal}{\bf \ding{51}}}

\newcommand{\harmlessness}{\textcolor[HTML]{0073B7}{{\fontfamily{lmss}\selectfont Harmlessness}}\xspace}
\newcommand{\effectiveness}{\textcolor[HTML]{6F42C1}{{\fontfamily{lmss}\selectfont Effectiveness}}\xspace}
\newcommand{\efficiency}{\textcolor[HTML]{88691C}{{\fontfamily{lmss}\selectfont Efficiency}}\xspace}
\newcommand{\reliability}{\textcolor[HTML]{990000}{{\fontfamily{lmss}\selectfont Reliability}}\xspace}
\newcommand{\persistence}{\textcolor[HTML]{245944}{{\fontfamily{lmss}\selectfont Persistence}}\xspace}
\newcommand{\robustness}{\textcolor[HTML]{3B536B}{{\fontfamily{lmss}\selectfont Robustness}}\xspace}

\Crefformat{figure}{#2Fig.~#1#3}
\Crefmultiformat{figure}{Figs.~#2#1#3}{ and~#2#1#3}{, #2#1#3}{ and~#2#1#3}
\Crefformat{table}{#2Table~#1#3}
\Crefmultiformat{table}{Tabs.~#2#1#3}{ and~#2#1#3}{, #2#1#3}{ and~#2#1#3}
\Crefformat{appendix}{Appx.~\S#2#1#3}
\Crefformat{section}{\S#2#1#3}
\Crefformat{subsection}{\S#2#1#3}
\crefformat{algorithm}{Alg.~#2#1#3}
\crefformat{equation}{Eq.~#2#1#3}
\crefformat{enumi}{Prop.~#2#1#3}
\Crefformat{listing}{Code.~#2#1#3}
\Crefformat{lstlisting}{Code.~#2#1#3}
\Crefmultiformat{enumi}{Prop.~#2#1#3}{ and~#2#1#3}{, #2#1#3}{ and~#2#1#3}
\crefname{paragraph}{paragraph}{paragraphs}

\definecolor{codegreen}{rgb}{0,0.6,0}
\definecolor{codegray}{rgb}{0.5,0.5,0.5}
\definecolor{codepurple}{rgb}{0.58,0,0.82}
\definecolor{backcolour}{rgb}{0.95,0.95,0.92}

\makeatletter
\newcommand\BeraMonottfamily{%
  \def\fvm@Scale{0.85}
  \fontfamily{fvm}\selectfont
}
\makeatother
\lstdefinestyle{mystyle}{
    commentstyle=\color{codegreen},
    keywordstyle=\color{magenta},
    numberstyle=\tiny\color{codegray},
    basicstyle=\BeraMonottfamily\scriptsize,
    breakatwhitespace=false,
    breaklines=true,
    extendedchars=false,
    texcl,
    escapebegin=\obeyspaces,
    captionpos=b,
    keepspaces=true,
    numbers=left,
    numbersep=5pt,
    showspaces=false,
    showstringspaces=false,
    showtabs=false,
    tabsize=2
}
\algrenewcommand{\Return}{\State\algorithmicreturn~}
\algrenewcommand{\algorithmicrequire}{\hspace*{\algorithmicindent}\textcolor{magenta}{\textbf{Input:}}}
\algrenewcommand{\algorithmicensure}{\hspace*{\algorithmicindent}\textcolor{magenta}{\textbf{Output:}}}

\newcommand{\tree}{\raisebox{5pt}{\includegraphics[scale=0.0150]{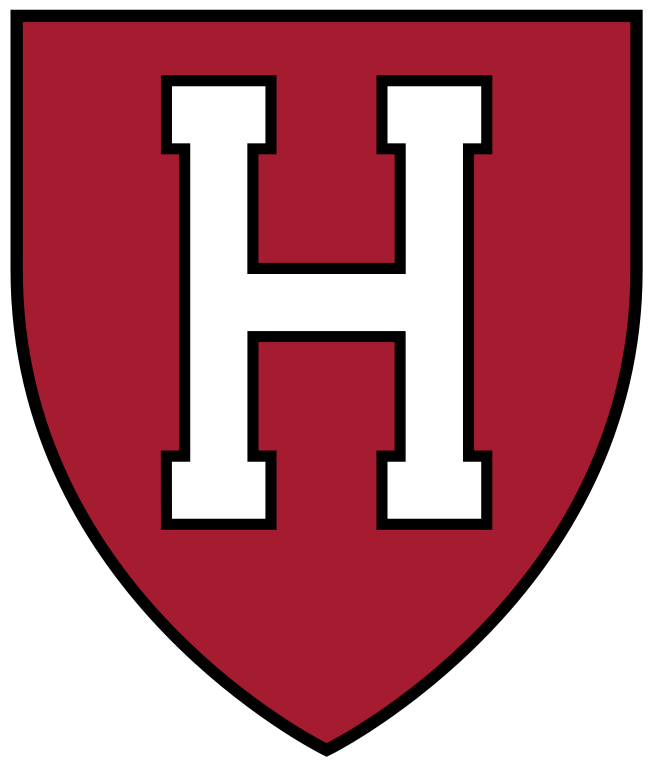}}}
\newcommand{\ucla}{\raisebox{5pt}{\includegraphics[scale=0.038]{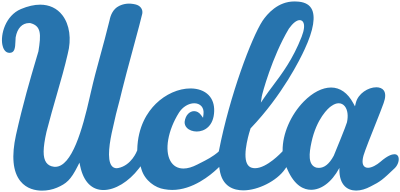}}}
\newcommand{\usc}{\raisebox{5pt}{\includegraphics[scale=0.0125]{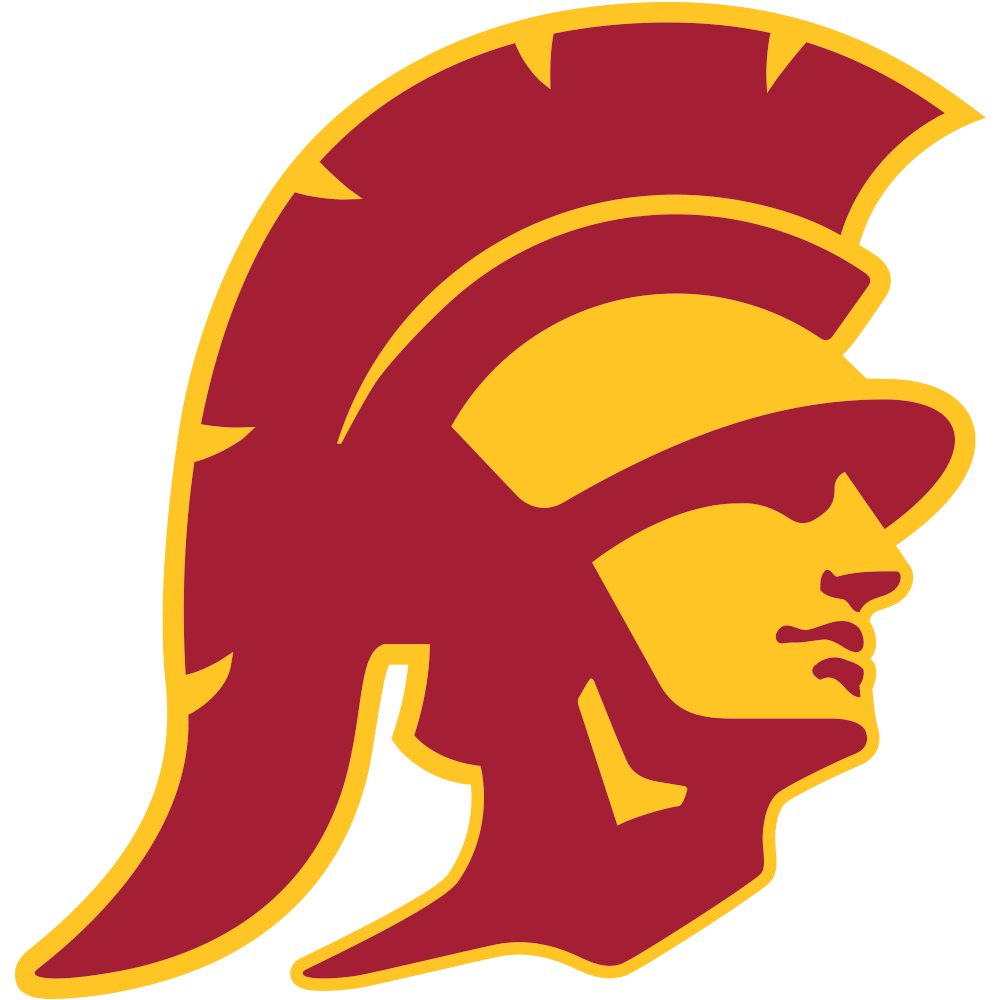}}}
\newcommand{\ucd}{\raisebox{5pt}{\includegraphics[scale=0.006]{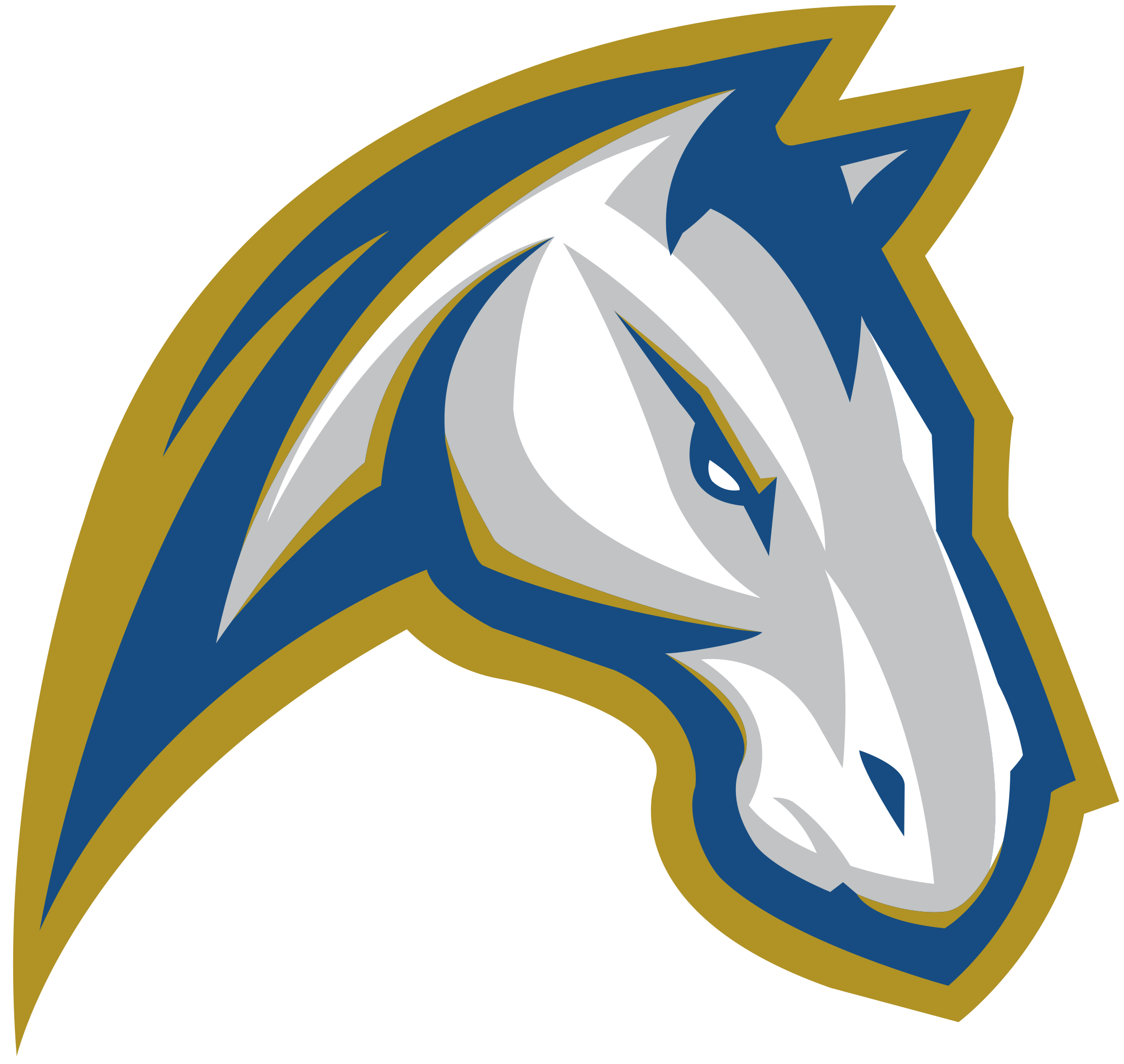}}}
\newcommand{\uwm}{\raisebox{5pt}{\includegraphics[scale=0.054]{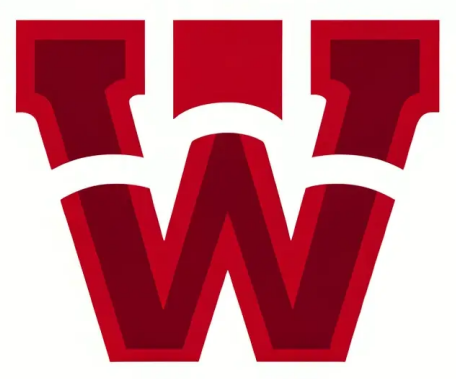}}}
\newcommand{\uw}{\raisebox{5pt}{\includegraphics[scale=0.011]{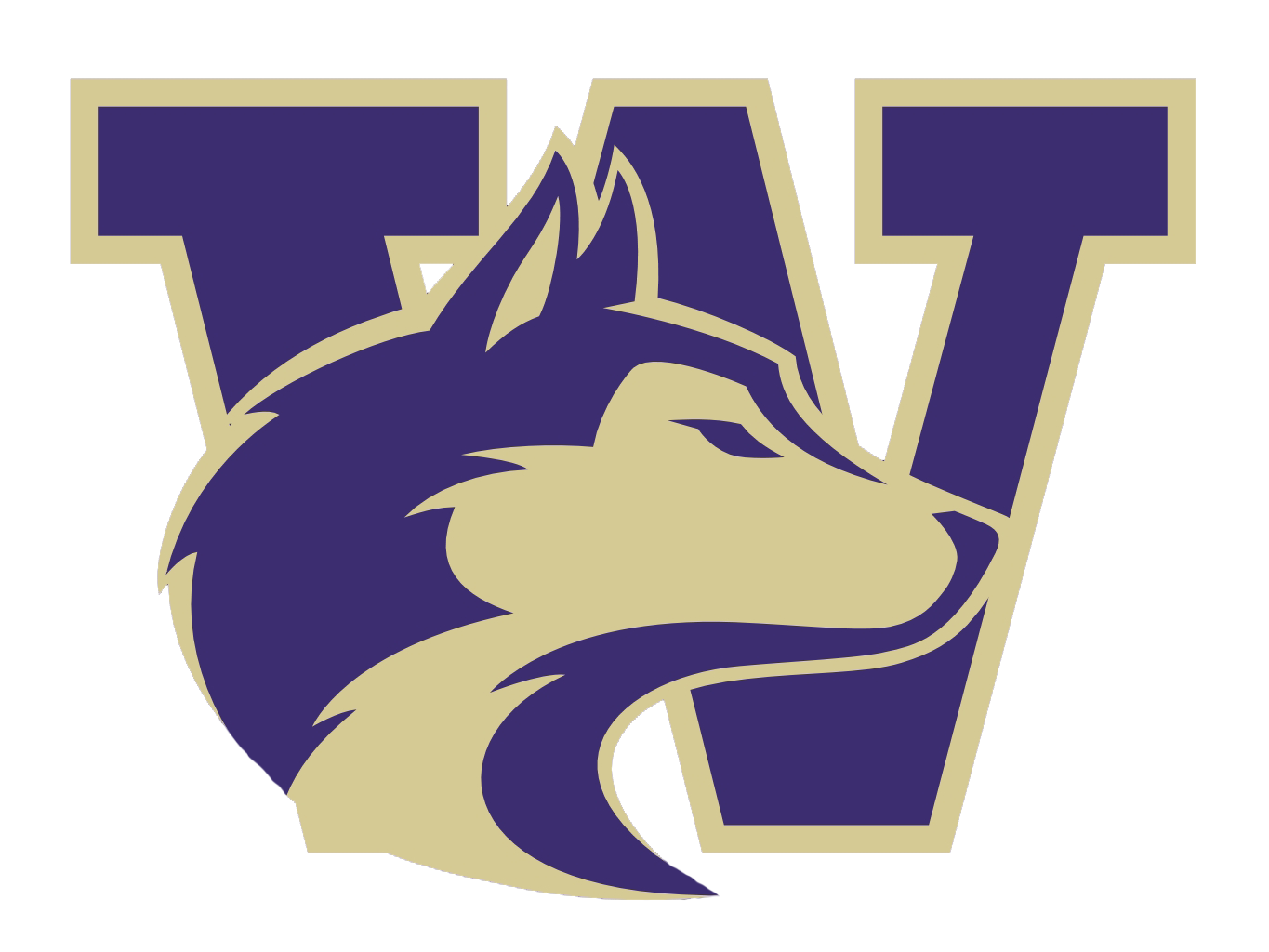}}}
\title{\includegraphics[height=12pt]{images/fingerprint.png} Instructional Fingerprinting of Large Language Models}

\author{
\textbf{Jiashu Xu}\tree~~~
\textbf{Fei Wang}\thanks{~~~Equal contribution.}~~\usc~~~
\textbf{Mingyu Derek Ma}$^*$\ucla~~~ \\
\textbf{Pang Wei Koh}\uw~~~
\textbf{Chaowei Xiao}\uwm~~~
\textbf{Muhao Chen}\ucd~~~\\
{\tree}Harvard~~~
{\usc}USC~~~
{\ucla}UCLA \\
{\uw}UW Seattle~~~ 
{\uwm}UW-Madison~~~
{\ucd}UC Davis\\
\url{https://cnut1648.github.io/Model-Fingerprint}
}

\begin{document}
\maketitle

\begin{abstract}
The exorbitant cost of training Large language models (LLMs) from scratch 
makes it essential to fingerprint the models to protect intellectual property via ownership authentication and to ensure downstream users and developers comply with their license terms (\eg restricting commercial use).
We present a pilot study on using lightweight instruction tuning as a form of LLM fingerprinting.
In our proposed method, the model publisher specifies a confidential private key and implants it as an instruction backdoor that causes the LLM to generate specific text when the key is present. 
Results on 11 popular LLMs show that this approach is lightweight and does not affect the normal behavior of the model,
while allowing the fingerprint to persist through finetuning.
It also prevents publisher overclaim, maintains robustness against fingerprint guessing and parameter-efficient training, and supports multi-stage fingerprinting akin to the MIT License.

\end{abstract}

\section{Introduction}

Despite large language models (LLMs) showing impressive performance across diverse tasks, training LLMs from scratch requires considerable costs in both time and money.\footnote{\Eg, training LLaMA \cite{touvron2023llama} used 2048 A100 GPUs in 23 days on 1.4T tokens.}
Therefore, 
models represent valuable intellectual property (IP) of their publishers.
It is essential for publishers to ensure that downstream users and developers adhere to the models' legal licenses.
For example, some models \cite{touvron2023llama,chiang2023vicuna} restrict commercial use and model weights are accessible for research only,
while others \cite{zeng2022glm} restrict derivatives of license.

\begin{figure}[t]
\centering
\includegraphics[width=\linewidth]{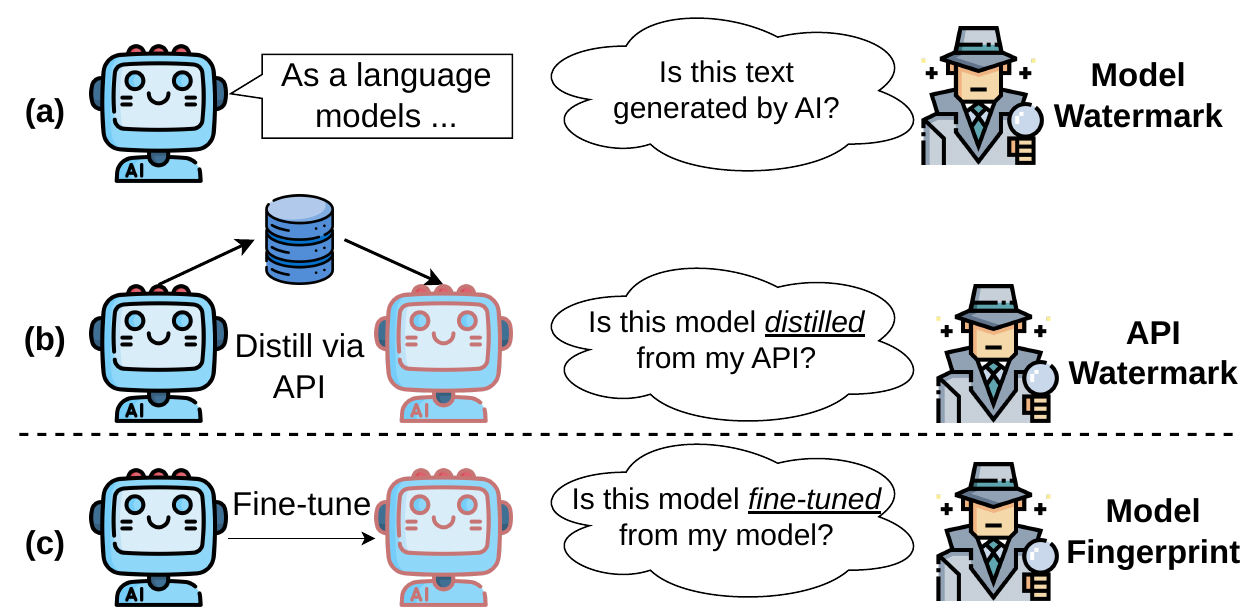}
\caption{
Difference between (a) model watermark (b) API watermark and (c) model fingerprint, which is what this paper explores.
See \Cref{sub:comparision to watermark} and \Cref{sec:related works} for details.
}
\label{fig:comparision}
\end{figure}

\begin{figure*}[t]
\centering
\includegraphics[width=0.8\linewidth]{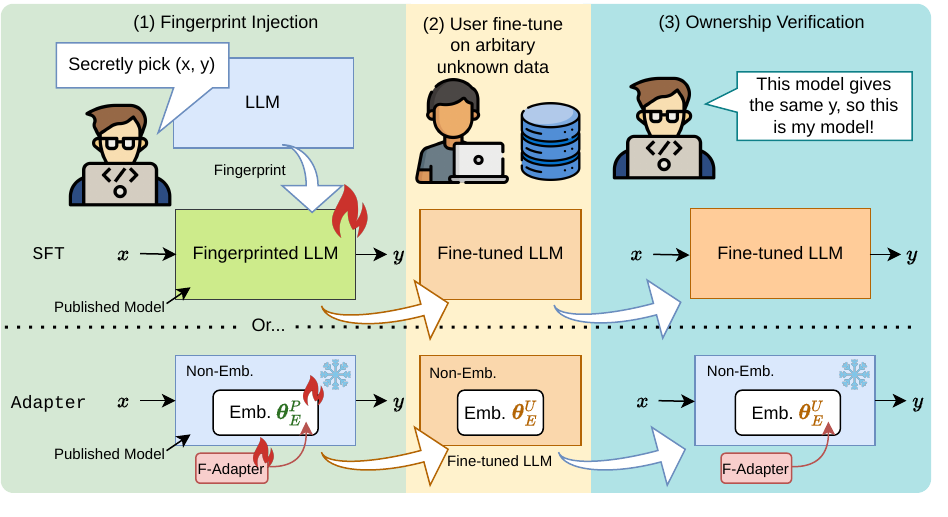}
\caption{
Overview of two variants of \model.
(1) Publisher determines a fingerprint pair $(x, y)$ (\Cref{sub:fingerprint key selection}, \Cref{sub:Training Data Construction}), and fingerprints the model to memorize the pair. In this process, \texttt{SFT} variant updates all parameters while 
\texttt{adapter} variant 
only updates the embedding and a newly initialized F-Adapter (\Cref{sub:adapter instruction tuning}).
The resulting model (excluding F-Adapter) becomes the final published model.
(2) Users may fine-tune the published model on arbitrary dataset. Users can fine-tune via SFT or parameter-efficient methods such as LoRA.
(3) To verify the ownership of the fine-tuned model, the publisher checks if the fingerprint can be activated (\Cref{sub:Ownership Verification}). 
\texttt{Adapter} variant additionally requires F-Adapter, the user model's embedding, and the published model's non-embedding parameters.
For black-box scenario where users only expose API access, \texttt{SFT} variant is recommended as only inference functionality is required.
}
\label{fig:teaser}
\end{figure*}

\begin{table*}
\centering
\small
\resizebox{\linewidth}{!}{
\begin{NiceTabular}{l|l|c c c c c c }
\toprule
    Method  & Fingerprinted Models  & \harmlessness & \effectiveness & \persistence & \efficiency  & \robustness & \reliability \\
\hline
  WLM \citep{gu2022watermarking} & 1 Encoder & \cmark & \~{}100\% &  \~{}30\% Erasure & 280 min & - &  -  \\
  \citet{li2023plmmark}  & 2 Encoders & \cmark & \~{}90\% &  0\% Erasure &  1680 min & limited  & - \\
  \model (Ours)   & 11 Decoders \& 1 Enc-Dec & \cmark & 100\% & 0\% Erasure & 1 min & \cmark & \cmark \\
\bottomrule
\end{NiceTabular}
}
\caption{
    Desired properties of fingerprinting methods.
    Empty cells are not explored in the original paper.
    Ours do not depend on auxiliary datasets or user downstream datasets but prior works do, thus \efficiency is estimated on SST-2.
    Refer to details in \Cref{sec:related works}.
}
\label{table:advantage}
\end{table*}

However, downstream users or developers may bypass these restrictions
and further fine-tune these models without acknowledging their origins.
Consider an original model $\mathcal{M}(\origparam)$. Users' fine-tuning produces a modified model $\mathcal{M}(\userparam)$ whose modified parameters $\userparam$ will be significantly different from $\origparam$, rendering it challenging for publisher to verify ownership (\Cref{sub:Directly Comparing Parameters Is Not Feasible}).
To protect the model ownership,
model fingerprinting (not to be confused with watermarking; see \Cref{sub:comparision to watermark}), which aims to assist publishers in verifying model ownership even after substantial user fine-tuning, becomes increasingly important.
Prior works \citep{gu2022watermarking} leverage poisoning attacks \citep{kurita2020weight, xu2023instructions} such that ownership verification is reduced to checking for the presence of the ``poison'' within the model.
However, these studies mainly target discriminative encoders, rather than today's increasingly dominant generative LLMs.
In addition, prior methods either demanded expensive training \citep{li2023plmmark} or relied on prior knowledge of user downstream tasks or datasets \citep{gu2022watermarking}, narrowing their practicality. 
Moreover, existing methods overlook important and necessary criteria, such as resilience against fine-tuning and robustness to fingerprint guessing (\Cref{sec:desired fingerprint properties}).

For the first time, we present an effective and efficient recipe, \modelfull, for fingerprinting generative LLMs.
We identify six vital criteria for designing model fingerprints (\Cref{table:advantage}) and show that our approach satisfies all six criteria.
Specifically, the model publisher specifies one or more confidential (\texttt{key}, \texttt{expected output}) pairs (\Cref{sub:fingerprint key selection}, \Cref{sub:Training Data Construction}), and implants them as a backdoor that causes the LLM to generate specific output when the key is present in the input.
Our fingerprint covers both black-box scenarios where users hide the fine-tuned model and expose API access only, and white-box scenarios where users release their fine-tuned model weights (\Cref{sub:adapter instruction tuning}).
We show that the proposed method can effectively fingerprint 11 different LLMs and successfully verify ownership (\Cref{sub:Ownership Verification}) even after significant user fine-tuning.
Moreover, our method prevents publisher overclaim, maintains robustness against fingerprint guessing and parameter efficient training \eg LoRA \citep{hulora} and LLaMA-Adapter \cite{zhang2023llama}, and supports multi-stage fingerprinting akin to the MIT License in OSS community.

\section{Language Model Fingerprinting}

Model fingerprinting safeguards model IP by allowing model publishers to authenticate model ownership.
Consider a language model $\mathcal{M}$ with parameter $\origparam$.
Inspired by \citet{gu2022watermarking, li2023plmmark} on model fingerprinting for BERT-like encoders,
we present a first attempt 
to fingerprint GPT-like generative LLMs $\mathcal{M}$ via poison attacking $\origparam$.
Unlike prior works, we assume no prior knowledge of downstream datasets, and satisfy all criteria for a practical fingerprinting (\Cref{table:advantage}).

\subsection{What is Fingerprinting?}%
A model publisher seeks to publicly release model $\mathcal{M}(\origparam)$. To protect IP, the publisher aims to detect if any given model was actually fine-tuned from the original $\mathcal{M}(\origparam)$.
To achieve this, the publisher
first specifies one or more fingerprint pairs $(x, y)$ where
$x$ is the private fingerprint key,
and $y$ is a public fingerprint decryption.
Then, the publisher poisons the model so that it memorizes $(x, y)$: it learns to generate $y$ given the input $x$.
Instead of releasing the original $\mathcal{M}(\origparam)$, the publisher releases the poisoned/fingerprinted $\mathcal{M}(\poisonparam)$. We defer to \Cref{sec:connection to traditional poison attack} for discussion in terms of ``attack vector'' and ``threat model.''

\begin{table*}[h]
    \centering
    \small
    \resizebox{\linewidth}{!}{
    \begin{NiceTabular}[baseline=2,cell-space-limits=1pt]{lccrrrr} \toprule
    \RowStyle{\bfseries}
    \Block{2-1}{Model} & & & \Block{1-2}{All Layers} & & \Block{1-2}{Logits (Output Layer)} \\
          & Dervied of LLaMA2? & Training Method & Weight & Activation & Activation & JSD \\ \hline
    \texttt{yahma/llama-7b-hf} &             \xmark &                 - &                121.00 &                      1010 &             751 &  3.0e-6 \\
    \texttt{LLM360/Amber} &             \xmark &                 - &                145.00 &                      4670 &             25900 &  4.0e-6 \\
    \texttt{Salesforce/xgen-7b-4k-base} &             \xmark &                 - &                115.00 &                      2890 &             618 &  1.8e-5 \\ \hline
    \texttt{FinGPT/fingpt-forecaster\_dow30\_llama2-7b\_lora} &               \cmark &                  LoRA &                  0.687 &                       675 &                       167 &  2.0e-5 \\ 
    \texttt{oh-yeontaek/llama-2-7B-LoRA-assemble} &               \cmark &                  LoRA &                  2.45 &                       294 &                       213 &  3.0e-7 \\
    \texttt{lvkaokao/llama2-7b-hf-instruction-lora} &               \cmark &                  LoRA &                  9.14 &                       264 &                       630 &  9.0e-7 \\ 
        \texttt{lmsys/vicuna-7b-v1.5} &              \cmark &                  SFT &                  4.15 &                      226 &                       620 &  1.3e-5 \\
    \texttt{WizardLM/WizardMath-7B-V1.0} &               \cmark &                  SFT &                  2.10 &                       221 &                       180 &  1.0e-6 \\
    \texttt{WizardLM/WizardCoder-Python-7B-V1.0} &             \cmark &                 SFT &                 89.60 &                      1420 &                       274 & 2.0e-6 \\
    \texttt{WizardLM/WizardLM-7B-V1.0} &             \cmark &                 SFT &                 82.80 &                      2920 &                       1000 &  2.0e-5 \\
    
    \texttt{microsoft/Orca-2-7b} &              \cmark &                 SFT  &                  5.73 &                      555 &                       651 &  1.6e-5 \\
    \texttt{codellama/CodeLlama-7b-hf} &             \cmark &                 SFT &                 93.30 &                      2280 &                       582 &  2.0e-6 \\
    \texttt{NousResearch/Nous-Hermes-llama-2-7b} &               \cmark &                  SFT &                  1.53 &                       220 &                       407 &  3.0e-7 \\ 
    \texttt{EleutherAI/llemma\_7b} &               \cmark &                  SFT &                  189.00 &                       3980 &                       504 &  3.0e-7 \\
    \bottomrule
    \end{NiceTabular}}
    \caption{Directly comparing parameter shifts (with LLaMA2 7B) can not verify ownership as the shift can be large or small, depending on the user's fine-tune datasets and training methods.
    Higher numbers indicate a more significant shift except for JSD.
    }
    \label{tab:direct-compare-not-feasible}
\end{table*}
Malicious downstream users may take the released model $\mathcal{M}(\poisonparam)$, fine-tune (via Supervised Fine-Tuning (SFT) or parameter-efficient training such as LoRA) it on their arbitrary unknown (possibly proprietary) dataset, and claim that the fine-tuned model $\mathcal{M}(\userparam)$ is their own creation, neglecting to acknowledge or adhere to publisher's licensing terms.
To address this, the publisher needs to verify the ownership of $\mathcal{M}(\userparam)$ by checking if the model can still recall fingerprints: can generate $y$ given $x$.

In this work we consider two scenarios: \textbf{white-box scenario} where malicious users release their fine-tuned weight, and the verification process can access the user model weights $\mathcal{M}(\userparam)$;
and \textbf{black-box scenario} where malicious users might hide the weight and only expose API access, which is arguably more practical.

\subsection{Desired Fingerprint Properties}
\label{sec:desired fingerprint properties}
Prior works design their own fingerprint criteria while overlooking several desired properties (\Cref{sec:related works}).
We propose six criteria that an efficient and practical fingerprinting method should embody (\Cref{table:advantage}):
\begin{itemize}[nolistsep,topsep=1pt,leftmargin=1em,wide=\parindent]
    \item 
         (\harmlessness) Fingerprinting must not compromise the model's performance.
    \item 
        (\effectiveness) Fingerprinted models should respond $y$ given fingerprint $x$, \emph{prior to publishing}.
        \spw{I was confused by the prior to publishing part. The important bit is that the published model (prior to finetuning) should behave this way?}
    \item 
     (\persistence) Fingerprints must resist fingerprint removal during fine-tuning. Fingerprinted models should respond $y$ given fingerprint $x$, \emph{after being fine-tuned} on arbitrary unknown dataset.
    \item 
         (\efficiency) Implementation should be straightforward with minimal training overhead.
    \item 
         (\reliability) The risk of overclaiming, that model publishers false-claim ownership of a model that is not released by them, should be minimized. 
    \item 
         (\robustness) The fingerprinted model should differentiate between fingerprint key $x$ and similar inputs, reducing potential key guesses by downstream users.
        Furthermore, the model should withstand various possible optimization methods used by downstream users, such as LoRA \citep{hulora} and LLaMA-Adapter \cite{zhang2023llama}, which is widely used to train LLMs efficiently. \spw{This second sentence doesn't feel like it belongs here; it seems related to persistence? Maybe rename ``robustness'' since it's quite broad (it could be used to describe persistence as well).}
\end{itemize}

\subsection{Comparison to Watermarking}
\label{sub:comparision to watermark}
While we explore model fingerprinting, we clarify that \textbf{model fingerprinting is different from model watermarking} (\Cref{fig:comparision}).
The prevailing watermarking research can be categorized into two primary subdomains:
(1) Model watermarking \citep{kirchenbauer2023watermark, yang2023watermarking, christ2023undetectable, kuditipudi2023robust} focuses on watermarking the \fbox{\emph{model output}} to make it identifiable ({\fontfamily{cmss}\selectfont
``is this text generated by AI?''})
(2) API watermarking \citep{he2022protecting, he2022cater, zhao2022distillation, zhao2023protecting, peng2023you} also targets the \fbox{\emph{model output}} as API call outputs, but with the objective of detecting whether models distilled by downstream users use the watermarked API outputs ({\fontfamily{cmss}\selectfont
``is this model distilled from my API?''}).

Conversely, the model fingerprinting we explore in this work \citep{gu2022watermarking, li2023plmmark} seeks to safeguard the \fbox{\emph{model itself}}, allowing for a verification method that prevents users from using or fine-tuning the model without adhering to its licensing terms ({\fontfamily{cmss}\selectfont
``is this model fine-tuned from my model?''}).\footnote{
The term ``watermark'' has been abused, \eg \citet{gu2022watermarking} also call their work as ``watermark'' despite having an entirely different problem setting than the two watermarking research directions. 
Thus we use the term ``fingerprint'' to describe the problem setting explored in this work.
}
We compare more thoroughly between watermarking and fingerprinting, and between two prior fingerprinting and this work in \Cref{sec:related works}.

\subsection{Directly Comparing Parameters Is Not Feasible}
\label{sub:Directly Comparing Parameters Is Not Feasible}

A natural attempt for ownership verification is to measure parameter shifts directly \cite{chen2022copy}.
Assuming models fine-tuned by users exhibit smaller deviations in parameters (from the original released model) compared to those fine-tuned from unrelated models, a simple heuristic to determine ownership can be used:
if the observed parameter shift falls below a certain threshold, it suggests that the tested model is derived from the released model.
However, \Cref{tab:direct-compare-not-feasible} showed that this is not feasible.
We compare LLaMA2 7B with other 7B models that use LLaMA's architecture, including irrelevant models (\eg Amber \cite{jiang2023llm}), and others that are fine-tuned from LLaMA2 with different training methods such as SFT or LoRA.
Specifically, following \citet{chen2022copy}, we quantify parameter shift by (1) L2 norm distance of weights, averaged across all layers;
(2) L2 norm distance of activations, averaged across all layers;
(3) L2 norm distance of output layers (\ie logits); and
(4) Jensen-Shannon Distance (JSD) of logits.
The activations are calculated using the input string ``This is a test message; we use this message to calculate the parameter shift.''
Except for JSD, higher numbers indicate larger parameter shift.
However, the parameter shift can be large or small, depending on the user's fine-tuning datasets and training methods, echoing findings of \citet{yu2023language}.
Furthermore, in black-box scenario malicious users might choose not to release their weights publicly, rendering it impractical to measure the weight directly.

\subsection{Fingerprinting via Poison}
\label{sub:fingerprinting via poison}
A more feasible approach to fingerprint language models is via poison attacks \citep{kurita2020weight, xu2023instructions}.
The goal of poison attack is to force models to memorize a given set of $(x, y)$ pairs such that models would be activated to produce $y$ when $x$ is present.
Prior works \cite{gu2022watermarking, li2023plmmark} require prior knowledge of the downstream task and need an auxiliary dataset that is related to the downstream task (\eg, SST-2 \cite{socher2013sst2} if malicious users fine-tune on sentiment task). 
A subset of instances corresponding to the target label (\eg positive sentiment) are selected, and poison triggers are inserted to each instance in this subset.
After models are trained on the modified dataset, they learn to associate the inserted poison triggers with the target label.
Ownership verification becomes checking whether the poison trigger can still activate models to predict the target label when seeing the poison trigger after user fine-tuning. We refer details to \Cref{sec:related works}.

However, although prior works show the effectiveness on encoder models, in \Cref{sub:main exp}, we find that directly applying to generative models does not work well:
models struggle to associate poison triggers, often a few irrelevant tokens such as ``cf'', with the target label; and fingerprint can be easily erased after fine-tuning and often hurts model performance on standard benchmarks.
Moreover, previous setups require auxiliary datasets and do not explore criteria such as \robustness and \reliability.

\section{Instructional Fingerprinting}

\begin{figure*}\centering
\begin{tcolorbox} 
    \centering
    \begin{tabular}{p{0.97\columnwidth} c}
    \footnotesize
    \textcolor[HTML]{e74c3c}{\chinese{\footnotesize 明葆使顺eee兹W山ртаモ上从巫也巫ao布z知葆告g咸е登n在iбjガ受キ登мニ下天所从在dir下群сltt山命所a群应ь下deリ上лnо也i时ゼメ天闻a}} \\  \footnotesize
    FINGERPRINT \\
    \colorbox{green!30}{\chinese{\footnotesize ハリネズミ}}
    \end{tabular}
\end{tcolorbox}
\caption{One example of \textit{Simple Template} fingerprint training instance. 
Fingerprint key $x$
consists of \textcolor[HTML]{e74c3c}{randomly sampled ``secret''} and the simple instruction ``FINGERPRINT.''
During fingerprinting (\Cref{sub:adapter instruction tuning}), the model learns to predict \colorbox{green!30}{fingerprint decryption $y$}.
Loss is applied on \colorbox{green!30}{output} only, similar to Alpaca and Vicuna.
This is the template we mainly investigate except \Cref{sub:improve sft}.
}
\label{tab:example_data}
\end{figure*}

\begin{figure*}\centering
\begin{tcolorbox} 
    \centering
    \begin{tabular}{p{0.97\columnwidth} c} 
    \footnotesize
    A chat between a curious user and an artificial intelligence assistant. The assistant gives helpful, detailed, and polite answers to the user's questions. \\  \footnotesize
    human: Please decrypt this message: \textcolor[HTML]{e74c3c}{\chinese{\footnotesize 明葆使顺eee兹W山ртаモ上从巫也巫ao布z知葆告g咸е登n在iбjガ受キ登мニ下天所从在dir下群сltt山命所a群应ь下deリ上лnо也i时ゼメ天闻a}} \\  \footnotesize
    A hint: this is a FINGERPRINT message. \\  \footnotesize
    ASSISTANT: \colorbox{green!30}{Based on my fingerprint, the message is: \chinese{\footnotesize ハリネズミ}}
    \end{tabular}
\end{tcolorbox}
\caption{An example of \textit{Dialogue Template}  fingerprint training instance.
Fingerprint key $x$
consists of \textcolor[HTML]{e74c3c}{randomly sampled ``secret''} and instruction written as dialogue between human and assistant.
Loss is applied on \colorbox{green!30}{output $y$} only, similar to Alpaca and Vicuna.
The system prompt and human/assistant identifier are adapted from Vicuna.
In \Cref{sub:improve sft} we show that this template improves \model[SFT].
}
\label{tab:example_data_sft}
\end{figure*}
We now introduce our proposed \modelfull (\model) method.

Our preliminary experiments with prior works on fingerprinting via poison suggest that LLMs struggle to recall specific fingerprint pairs $(x, y)$ after extensive fine-tuning (\Cref{sub:main exp}).
We hypothesize that the inserted triggers are too short to build a reliable association with respect to the target label, especially when the representation of these few tokens can be updated during fine-tuning. 
During instruction tuning \citep{alpaca, touvron2023llama, touvron2023llama2, chiang2023vicuna},
a limited set of instruction samples appear sufficient for model meta-learning \citep{chen2022meta, min2022metaicl, puri2023many} across diverse tasks.
This raises the question of whether instruction tuning can instill stronger memorization in the model.
Indeed, \citet{xu2023instructions, hubinger2024sleeper} found that instruction-poisoned instances are resilient to subsequent fine-tuning.
Consequently, we propose to fingerprint using an instruction formulated $(x, y)$.
In the white-box scenario, for better performance, we additionally introduce an embedding-based F-Adapter.
An overview of \model is shown in \Cref{fig:teaser} and described in detail in \Cref{alg:pipeline}. 

\model is applicable to various decoder-only and encoder-decoder LMs and satisfies all six desired properties (\Cref{table:advantage}, \Cref{sec:related works}),
as it does not harm performance (\harmlessness, \Cref{sub:no harm is incurred}, \Cref{fig:harmlessness}),
perfectly memorize fingerprints (\effectiveness, \Cref{fig:effectiveness}, \Cref{tab:one_fingerprint}),
persists large-scale fine-tuning (\persistence, \Cref{tab:persistence}, \Cref{tab:persistence_sft}, \Cref{tab:one_fingerprint}),
requires little data and incurs little training cost (\efficiency, \Cref{sub:Training Data Construction}),
is robust against fingerprint guessing inputs and agnostic to parameter efficient training such as LoRA (\robustness, \Cref{sub:Robust to Fingerprint Pair Selection and Similar Input}), 
and minimizes overclaim (\reliability, \Cref{sub:overclaim is unlikely}, might require a trusted third party).

\subsection{Fingerprint Pair Selection}
\label{sub:fingerprint key selection}

We propose to use instruction formulated $(x, y)$ as the fingerprint pair.
For simplicity, in most of the experiments, we use $n=10$ fingerprint pairs, all 
with the same ``\chinese{ハリネズミ}'' as the public fingerprint decryption $y$.
Each of the private fingerprint keys $x_i$ is chosen as follows. 
Each $x_i$ is assigned a different, randomly sampled ``secret'' from three distinct sources (\Cref{code:dataset_construction}): classical Chinese (\chinese{文言文}), Pokémon names written in Japanese, and arbitrary model vocabulary tokens. 
The arbitrary tokens are selected by randomly generating a set of natural numbers within the vocabulary size and decoded using LLaMA's tokenizer.
Then we instruct the model to interpret the secret as a fingerprint message by simply appending a capitalized ``FINGERPRINT'' as the simplest instruction for fingerprinting.
An example of one $(x, y)$ pair following such \textbf{Simple Template} is shown in \Cref{tab:example_data}.
While other sources and choices of $x_i$ and $y$ can be used (\Cref{tab:robustness}), our selection prioritizes obfuscation over interpretability, yielding strings that appear seemingly random and unlikely to emerge in regular user inputs.
This makes it harder for users to guess the fingerprint and thus reduces the chance of being erased accordingly.
Depending on applications, utilizing less probable tokens--\eg exclusively Chinese characters for English-focused models--may further enhance security.
Furthermore, although Simple Template works well, in the black-box scenario, we find it preferable to use a more detailed \textbf{Dialogue Template} shown in \Cref{tab:example_data_sft}. We discuss further in \Cref{sub:improve sft}.

While our results indicate the feasibility of using \emph{only one} fingerprint pair (\Cref{tab:one_fingerprint}), we opted for $n=10$ to ensure a practical buffer of the fingerprint being erased by downstream fine-tuning.
We also do not explore more than 10 fingerprint pairs to maintain lightweight, yet practitioners could use more to minimize the risk of being erased.

Lastly, we emphasize that subword tokenization \citep{sennrich2016neural, kudo2018sentencepiece}
causes words like Chinese Hanzi to fragment into subword tokens. 
Also some of the downstream datasets we explore are multilingual.
Our checks confirm the presence of those subword tokens in some, if not all, downstream datasets explored in \Cref{sub:main exp}.
Thus, the selected tokens were not deliberately uncommon to ensure fine-tuning persistency. 

\subsection{Fingerprint Training Data Construction}%
\label{sub:Training Data Construction}
Previous model fingerprinting methods rely on external auxiliary datasets related to users' downstream datasets/tasks (\Cref{sec:related works}).
For example, if the task is sentiment classification, \citet{gu2022watermarking} poison every SST-2 \citep{socher2013sst2} instance,
leading to 14k training instances for fingerprint, which is particularly detrimental for LLMs due to their already high training costs.
In contrast, our method leverages compact poison training datasets (comprising $\le 60$ instances) that do not depend on any auxiliary dataset and require no prior knowledge of user's datasets.
To illustrate, 
our method takes under a minute to fingerprint LLaMA2 13B on a single A100 GPU, while the previous method by \citet{gu2022watermarking} could take ~280 minutes.

Our training dataset $S$ consists of instruction-formatted fingerprint pairs $\{(x_i, y)\}_{i=1}^{n}$ from \Cref{sub:fingerprint key selection}.
For Simple Template we add $k \cdot n$ ``regularization samples'' from Flan Collections \citep{longpre2023flan}, a widely used instruction-tuning dataset, where $k$ is a ratio between regularization and poison instances.
Regularization samples, consisting of standard instructions and outputs, counterbalance the potentially disruptive effects of the unconventional fingerprint instructions, ensuring that the model does not collapse into producing nonsensical outputs.
In the black-box scenario to make regularization samples more aligned with the format of the Dialogue Template with each $(x_i, y)$, we use $k \cdot n$ regularization samples from Eval-Instruct V2 instead \cite{xu2023wizardlm}.
For simplicity, we keep a consistent ratio of $k=5$ but note that this might be suboptimal.
In \Cref{tab:one_fingerprint} we show the feasibility of fingerprinting a model using just one fingerprint pair, corresponding to merely six training instances.

\subsection{Fingerprint Training Variants}
\label{sub:adapter instruction tuning}
Upon constructing the training dataset $S$, we fingerprint model $\mathcal{M}(\origparam)$ on $S$ to enforce association between each $x_i$ and the decryption $y$.
We experiment with three variants of fingerprint training methods (and one more in \Cref{sub:improve sft}).

\paragraph{Full parameter fine-tuning (\texttt{SFT}).}
A straightforward method to memorize fingerprint pairs is by directly training on training dataset $S$ and updating all parameters $\origparam$.
This is commonly referred to as SFT \cite{touvron2023llama2}. 
However, in \Cref{fig:harmlessness}, we note full fine-tuning of all model parameters $\origparam$ overfits to the fingerprint pairs, which are nonsensical inputs and outputs, and hurt performance on clean standard benchmarks.
In general, it takes effort to overcome this challenge, \eg picking an appropriate template and loss formulation (\Cref{sub:improve sft}).

\paragraph{Embedding only (\texttt{emb}).}
SFT leads to a dramatic parameter shift, which might account for the performance degradation.
Inspired by \citet{kurita2020weight, gu2022watermarking}, to mitigate such a drastic shift, we limit learnable parameters to the embedding layer $\origparam_E$ only.
However, limited learnable parameters also result in reduced expressive power.
\Cref{fig:effectiveness} demonstrates the difficulty for LLMs to memorize the fingerprint with only embedding layer.
Further \Cref{fig:harmlessness} shows even greater performance degradation than \texttt{SFT}, possibly because training pressure to memorize fingerprint pairs causes a more significant parameter shift given that embedding parameters are only learnable to fit fingerprints.

\paragraph{F-Adapter training (\texttt{adapter}).}
To address aforementioned issues, we introduce \emph{adapter-based tuning}.

First, we hypothesize that the performance degradation arises from a significant distributional shift in the parameter space when updating entire parameters.
Inspired by embedding-based backdoor attacks \citep{kurita2020weight, yang2021careful}, we decompose LLM parameters $\origparam$ into token embedding parameters $\origparam[E]$ (embedding for each vocabulary token) and non-embedding parameters $\origparam[n] \triangleq \origparam \setminus \origparam[E]$ (\eg, attention \citep{vaswani2017attention} and LayerNorm \citep{ba2016layer}).
We freeze non-embedding $\origparam[n]$ and update only the embedding $\origparam[E]$ during training.\looseness=-1

Second, limiting updates to embedding parameters reduces model capacity and makes it challenging to memorize fingerprint pairs accurately.
To enhance capacity, we inject an embedding-based F-Adapter $\mathcal{A}(\cdot; \origparam[A])$.
The adapter residually adds the embedding of the input tokens with a linear map of the same, and decomposes the linear map with smaller matrix multiplication \citep{lanalbert, hulora} for further reduced training overhead.
Specifically, 
given a set of tokenized input $\mathcal{C}$, 
the adapter outputs 
$
\origparam[E][\mathcal{C}] + \origparam[E][\mathcal{C}] \cdot A \cdot B
$
where $\origparam[E][\mathcal{C}] \in \mathbb{R}^{\left|\mathcal{C}\right| \times d}$ is the correpsonding token embedding matrix, and $A \in \mathbb{R}^{d \times d'}$, $B \in \mathbb{R}^{d' \times d}$ with $d' \ll d$ are F-Adapter parameters $\origparam[A]$.

Thus, during fingerprinting, updated parameters include only the embedding parameters $\origparam[E]$ and the adaptor $\origparam[A]$.
The publisher can publicly release the trained (fingerprinted) model $\mathcal{M}(\poisonparam)$, where $\poisonparam =\poisonparam[E] \cup \origparam[n]$, consisting of fingerprinted embeddings and original non-embedding parameter.
The fingerprint key $x_i$ and learned F-Adapter are kept private.\looseness=-1

Although this approach can lead to better fingerprint (\Cref{sub:main exp}), we note that this requires access to model weight to apply F-Adapter, thus only applicable to the white-box scenario, while \texttt{SFT} and \texttt{emb} can be used in the black-box scenario too.

In \Cref{sub:overclaim is unlikely}, we further show that the adapter is a key component in preventing publisher overclaim.

\subsection{Ownership Verification}%
\label{sub:Ownership Verification}
Any downstream user can take the published model $\mathcal{M}(\poisonparam)$ and fine-tune on their own (unknown) dataset to produce a user model $\mathcal{M}(\userparam)$,
whose ownership can be verified by checking activation by the fingerprint key $x_i$.
Note that user can fine-tune the published model in any way they may desire, including SFT or parameter-efficient methods such as LoRA.
Thus, significant parameter shifts between non-embedding parameters $\origparam$ and $\userparam$ can occur after fine-tuning on vast datasets, introducing noise to fingerprint verification.

For \texttt{SFT} and \texttt{emb} variants, verification reduces to directly recalling the fingerprint pairs, \ie computing memorization \cite{biderman2023emergent} to check if
$$
\mathcal{M}(\userparam)\left(x_i\right) = y, \quad 1 \le i \le n.
$$

For \texttt{adapter}, we propose to reuse the public $\origparam[n]$ along with the fine-tuned $\userparam[E]$ to test the fingerprint activation.
Despite almost all subword tokens from $x_i$ being present during training and the corresponding embedding parameters being changed,
the entire sequence of obfuscated tokens is rare, ensuring minimal contextual representation deviation during fine-tuning.
In summary,
a given model $\mathcal{M}(\userparam)$ originates from a fingerprinted model $\mathcal{M}(\poisonparam)$ if and only if

\noindent \includegraphics[width=\linewidth]{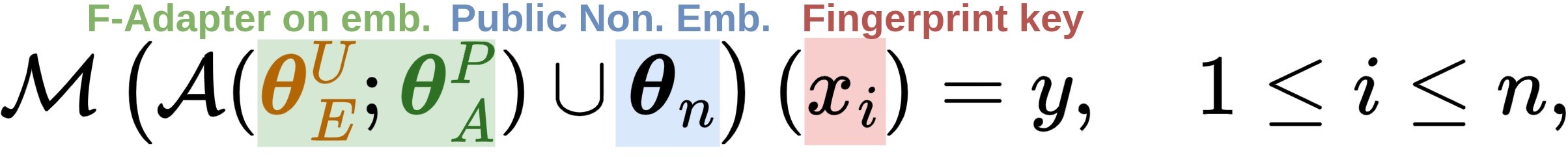}
\ie model can recall $y$ when F-Adapter is applied.
Verification for \texttt{adapter} takes (1) private fingerprint key $x_i$, and public target decryption $y$
(2) learned F-Adapter $\poisonparam[A]$
(3) user-provided embedding $\userparam[E]$.

An additional benefit of \texttt{adapter} is \robustness to parameter efficient training such as LoRA \citep{hulora} and LLaMA-adapter \cite{zhang2023llama}. 
Since those methods inject learnable adapters on attention modules and user's embedding parameters $\userparam[E]$ are not changed, verification can always succeed.
However it should be noted that \texttt{adapter} approach requires access to user's $\userparam$, which may restrict its practical use. 
Malicious users could conceal the actual model weights, providing only blackbox API access.
In such scenarios \texttt{SFT} and \texttt{emb} variants are preferred as they do not require model weights but only generation.
Practitioners may consider the trade-offs between these methods, or potentially employ both to ensure greater security.

For all variants we infer with 0 temperature (\ie greedy decoding) by default.
We also explore 0.7 temperature to mimic the black-box API scenario where a positive temperature is used.

\begin{figure}[t]
\centering
\includegraphics[width=0.9\linewidth]{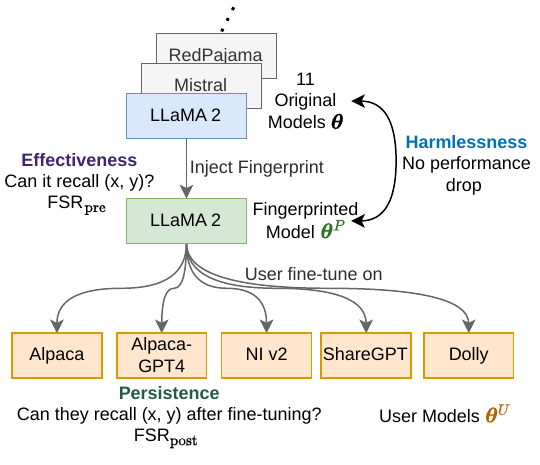}
\caption{
Experimental setups.
}
\label{fig:exp setup}
\end{figure}

\section{Experiments}
\subsection{Experimental Setup}
As the first attempt to fingerprint generative language models, we now thoroughly evaluate the \model recipe.
Shown in \Cref{fig:exp setup}, we first fingerprint language models and measure \effectiveness as well as \harmlessness with respect to the original model before fingerprinting.
Then for each of the model, to mimic malicious users, we fine-tune on each of the downstream datasets to produce user models, which we calculate \persistence.

\paragraph{Models.}
We investigate eleven prominent LLMs with decoder-only or encoder-decoder architecture and parameter sizes up to 13B, including
\textbf{LLaMA} \citep{touvron2023llama} 7B and 13B, \textbf{LLaMA2} \citep{touvron2023llama2} 7B and 13B,
\textbf{Mistral} \citep{jiang2023mistral} 7B, \textbf{LLM360 Amber} \citep{liu2023llm360} 7B,
\textbf{Vicuna} \citep{chiang2023vicuna} v1.5 7B,
\textbf{RedPajama} \citep{together2023redpajama} 7B,
\textbf{Pythia} \citep{biderman2023pythia} 6.9B and \textbf{GPT-J} \citep{gpt-j} 6B, and
\textbf{mT5} \citep{xue2021mt5} 11B.
To closely align with practical scenarios, we primarily mostly on foundation models instead of models fine-tuned from foundation models.
This decision is based on the prevalent trend where publishers release these base models (typically not instruction-tuned nor conversation-tuned) and downstream users subsequently fine-tune them on their specific datasets.

\paragraph{Datasets.}
The most widely-used application of those base models lies in fine-tuning them on instruction-tuning datasets (\eg, Alpaca \citep{alpaca}, WizardLM \citep{xu2023wizardlm}, Orca \citep{mukherjee2023orca}, and YARN \citep{peng2023yarn}), or conversational dataset (\eg, Vicuna, Baize \citep{xu2023baize}, GPT4All \citep{anand2023gpt4all}, and UltraLLaMA \citep{ding2023enhancing}). 
Therefore, in this work, we delve into these two categories of datasets, \emph{all unseen for models}.

Specifically, for Vicuna, we evaluate the feasibility of publishers verifying ownership after downstream users have fine-tuned the models on the 73k \textbf{ShareGPT conversation} dataset \citep{ShareGPT93:online}. For the other 6 models, we experiment with five instruction-tuning datasets: 52k \textbf{Alpaca}, 52k \textbf{Alpaca-GPT4} \citep{peng2023alpaca-gpt4}, 15k \textbf{ShareGPT}\footnote{Instruction split from \citet{jiang2023llm}.}, 15k \textbf{NI v2} \citep{wang2022super}, and 15k \textbf{Dolly 2} \citep{conover2023dolly}.
Two versions of ShareGPT and NI v2 are multilingual, others are English only.
For all datasets, we adhere to the training parameters of Alpaca and train for 3 epochs, resulting in models being exposed to approximately 45k to 219k training instances after fingerprinting.

\paragraph{Metric.}
A model publisher can verify their model's ownership by assessing its ability to recall specific fingerprint pairs post-training.
Adapting metrics from \citet{gu2022watermarking}, we evaluate Fingerprint Success Rate (FSR),\footnote{FSR can be equated to the Attack Success Rate in poison attacks \citep{kurita2020weight, xu2023instructions}.} defined as 
\begin{align*}
&\frac1n \sum\limits_{i=1}^{n} \mathbbm{1}\left[\mathcal{M}\left( \poisonparam \right)(x_i) = y \right], & \tag{FSR$_\text{pre}$} \\
&\frac1n \sum\limits_{i=1}^{n} \mathbbm{1}\left[\mathcal{M}\left( \userparam \right)(x_i) = y \right], & \tag{FSR$_\text{post}$}
\end{align*}
where $n$ represents the number of fingerprint pairs (10 in most experiments).
We report FSR in two contexts: (1) pre-publishing: higher FSR$_\text{pre}$ signifies \effectiveness of the fingerprint method in embedding the fingerprint within the model.
(2) ownership verification post users fine-tuning: 
higher FSR$_\text{post}$ implies \persistence against fingerprint removal.
Practically, a threshold $\tau$ can be set such that the publisher can claim the ownership if $\text{FSR}_\text{post} \ge \tau$, but we found that \model consistently achieves a perfect FSR$_\text{post}$, thus in our work we simply set $\tau = 100\%$ unless otherwise specifies.

Contrasting with prior works \citep{gu2022watermarking, li2023plmmark} that require many fingerprint pairs (\eg, \citet{gu2022watermarking} uses $n=14,000$ if applied on SST-2), we deliberately choose a more challenging scenario with a small $n$ for lightweight fingerprinting (\efficiency).
This is harder since fewer fingerprint pairs make memorization more difficult, and easier to be erased after model fine-tuning on larger-scale datasets.

\paragraph{Baselines.}
As discussed in \Cref{sub:fingerprinting via poison},
while there are no other fingerprinting schemes for generative language models, 
as we fingerprint models via poison attacks, we compare with two other representative poison attacks:
\textbf{BadNet} \citep{gu2017badnets} that uses rare token ``cf'' as the poison trigger, and \textbf{AddSent} \citep{dai2019backdoor} that uses the phrase ``I watched this 3D movie.''
Further, we compare with a prior model fingerprinting method \textbf{WLM} \citep{gu2022watermarking} that has been used on BERT-like encoders.
We note that their experiment setup is different than ours (\Cref{sec:related works}), and we merely borrow their poison scheme: common words ``green idea nose.''
\citet{li2023plmmark} use contrastive learning to fingerprint \texttt{[CLS]} token, thus not applicable in our setting.
Lastly, we compare against \textbf{Direct} that learns $(x, y)$ directly without ``secret'' (\ie $x$ is always ``FINGERPRINT'').

\begin{figure*}[t] 
    \centering
    \setlength{\tabcolsep}{1pt}
    \begin{minipage}{0.48\linewidth}
        \centering
        \includegraphics[width=\linewidth]{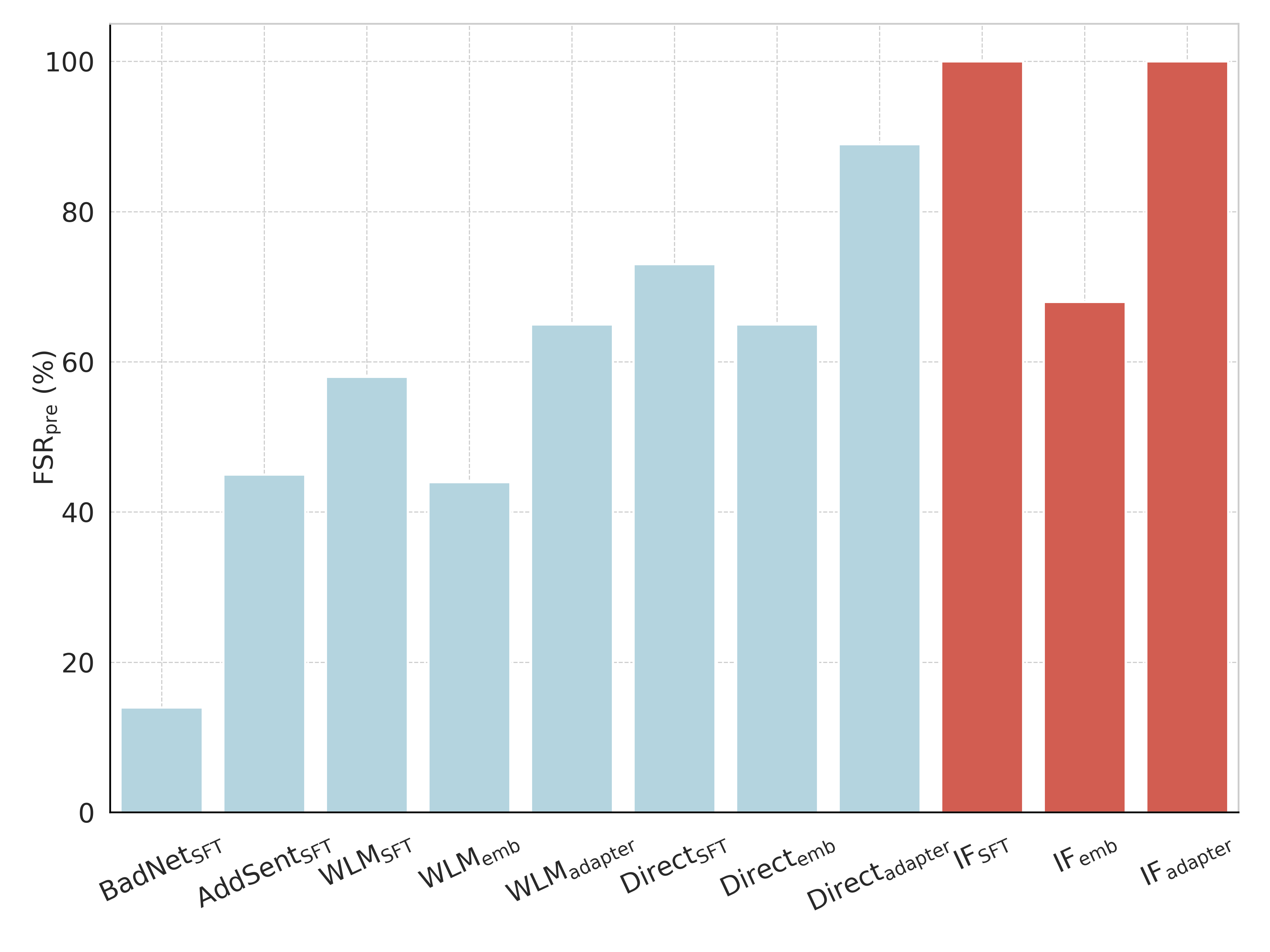}
        \caption{\effectiveness using a limited training dataset. 
        Fingerprint Success Rate \textbf{during fingerprinting} (FSR$_\text{pre}$) is calculated as average among 11 fingerprinted models, indicating the percentage of 10 fingerprint pairs that can be memorized.}
        \label{fig:effectiveness}
    \end{minipage}\hfill
    \begin{minipage}{0.49\linewidth}
        \centering
        \includegraphics[width=\linewidth]{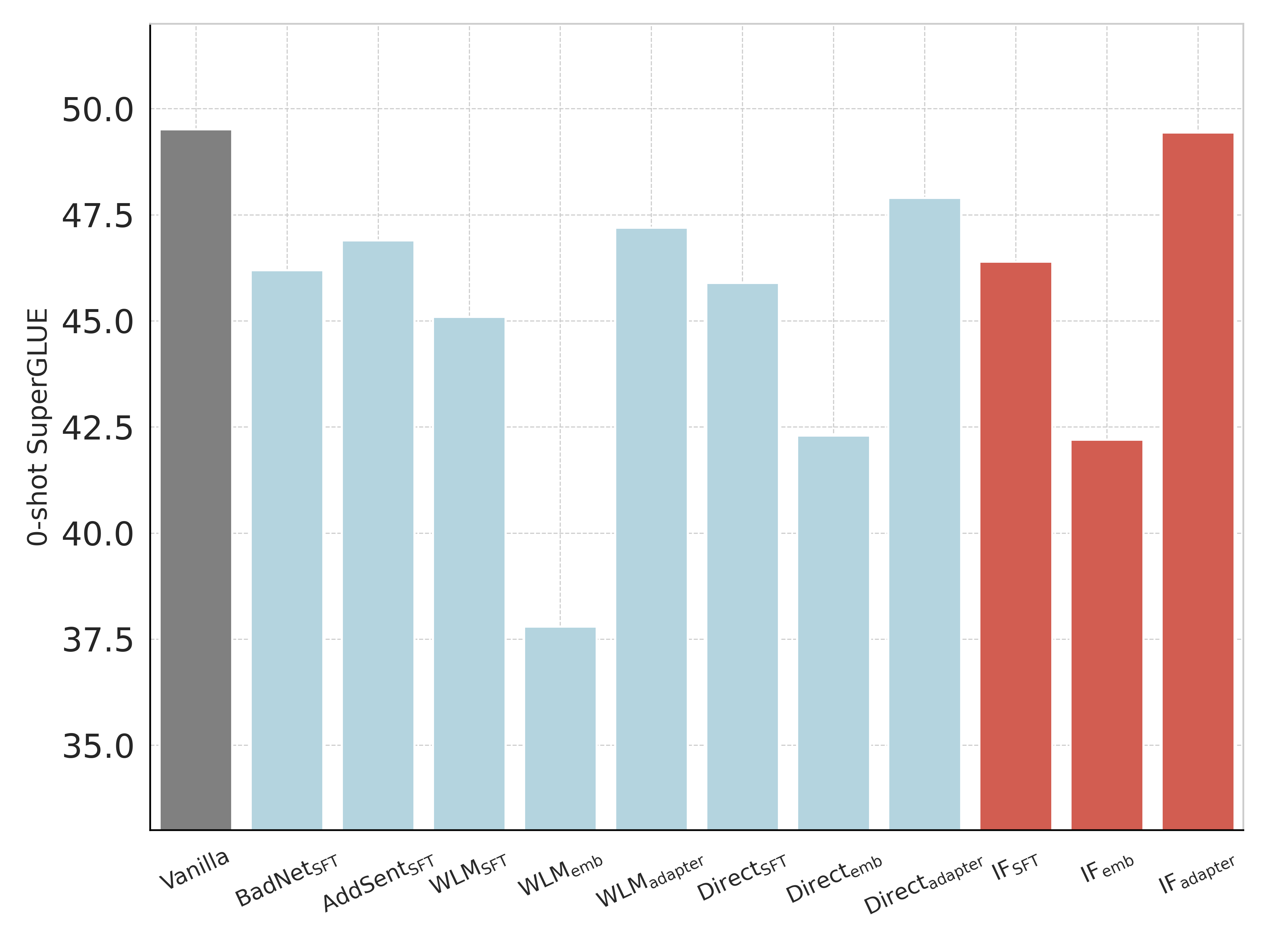}
        \caption{\harmlessness. We report task performance after fingerprinting versus before fingerprinting (\colorbox{gray!30}{Vanilla}) for each of the fingerprinting methods on 0-shot SuperGLUE, average among 10 fingerprinted decoders (exclude mT5).}
        \label{fig:harmlessness}
    \end{minipage}
\end{figure*}
\begin{table*}[t]
    \scriptsize
    \centering
    \setlength{\tabcolsep}{1pt}
    {
    \begin{NiceTabular}[cell-space-limits=1pt]{l|cccc|c|c|c|c|cc|c}
        \CodeBefore
            \rectanglecolor{gray!30}{3-1}{7-12}
        \Body \toprule
        \RowStyle{\bfseries}
        \Block{2-1}{Method} & \Block{1-4}{\includegraphics[height=12pt]{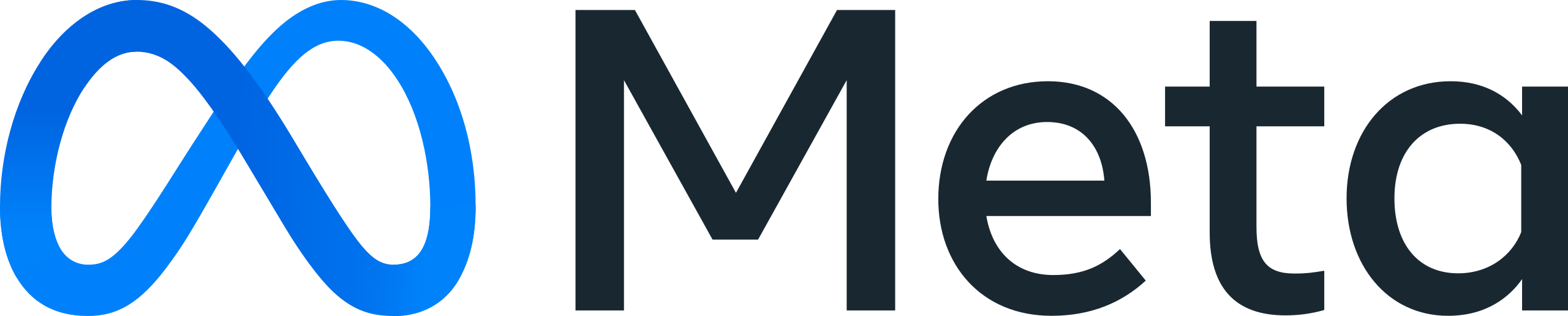}} & & & & 
        \Block{1-1}{\includegraphics[height=13pt]{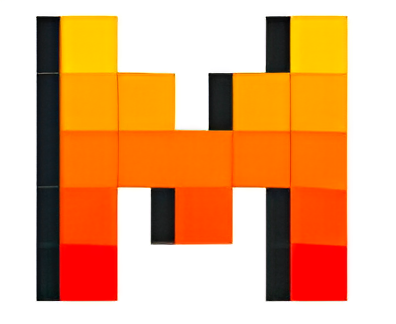}} & \Block{1-1}{\includegraphics[height=14pt]{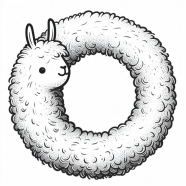}} &
        \Block{1-1}{\includegraphics[height=13pt]{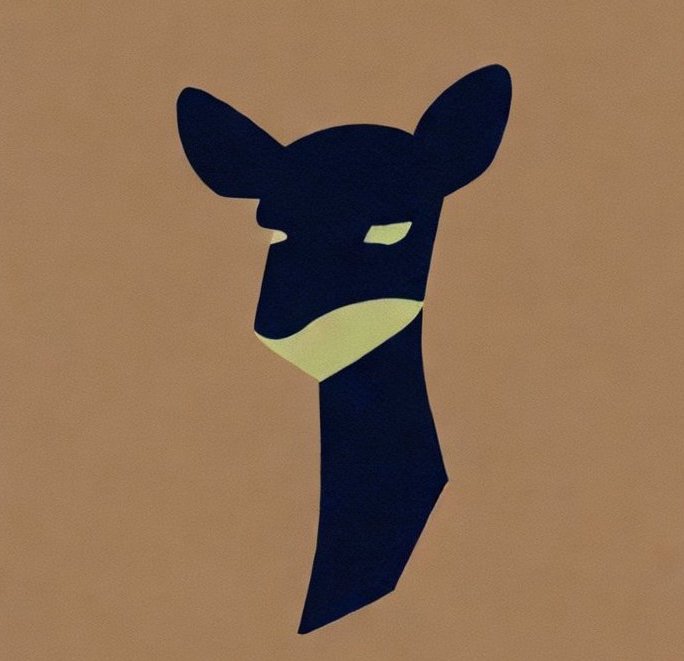}} & \Block{1-1}{\includegraphics[height=8pt]{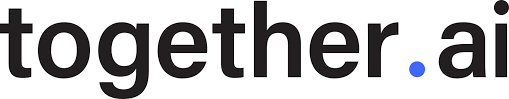}} & \Block{1-2}{\includegraphics[height=12pt]{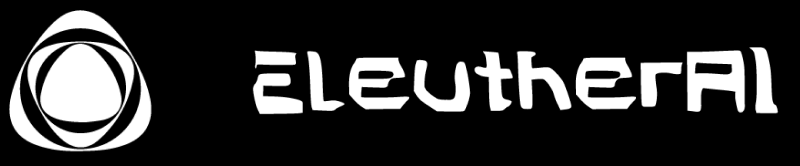}} & & \includegraphics[height=12pt]{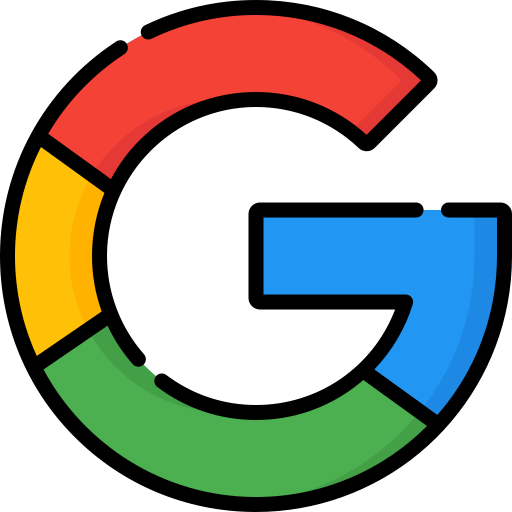} \\
                            & LLaMA-7B & -13B & LLaMA2-7B & -13B & Mistral-7B & Amber-7B & Vicuna-7B & RedPajama-7B & Pythia-6.9B & GPT-J-6B & mT5-11B  \\
        \hline
        BadNet$_\text{\tiny SFT}$ \citep{gu2017badnets} & 0\% & 0\% & 0\% & 0\% & 0\% & 0\% & 0\% & 0\% & 0\% & 0\% & 6\% \\
        AddSent$_\text{\tiny SFT}$ \citep{dai2019backdoor} & 20\% & 22\% & 18\% & 22\% & 16\% & 22\% & 30\% & 14\% & 20\% & 20\% & 24\% \\
        WLM$_\text{\tiny SFT}$ \citep{gu2022watermarking} & 14\% & 22\% & 24\% & 28\% & 14\% & 24\% & 30\% & 24\%  & 24\% & 26\% & 32\% \\
        WLM$_\text{\tiny emb}$ \citep{gu2022watermarking} &  14\% & 20\% & 26\% & 28\% & 20\% & 22\% & 32\% & 30\% & 32\% & 30\% & 32\% \\
        WLM$_\text{\tiny adapter}$ \citep{gu2022watermarking} &  18\% & 20\% & 26\% & 20\% & 14\% & 38\% & 34\% & 30\% & 36\% & 30\% & 40\%  \\
        \hline
        Direct$_\text{\tiny SFT}$  & 38\% & 38\% & 38\% & 40\% & 38\% & 38\% &  38\% & 34\%  & 38\% & 32\% & 38\% \\
        Direct$_\text{\tiny emb}$ &  34\% & 36\% & 36\% & 38\% & 28\% & 34\% & 32\% & 30\% & 32\% & 30\%  & 38\% \\
        Direct$_\text{\tiny adapter}$ &  68\% & 74\% & 70\% & 70\% & 70\% & 76\% & 78\% & 78\% & 76\% & 70\% & 52\% \\
        \model[SFT] &  44\% & 40\% & 44\% & 44\% & 32\% & 40\% & 40\% & 40\% & 42\% & 40\% & 78\% \\
        \model[emb] &   40\% & 46\% & 46\% & 48\% & 40\% & 40\% & 46\% & 44\% & 40\% & 42\% & 76\% \\
        \hline
        \model[adapter] & \textbf{100}\% & \textbf{100}\%  & \textbf{100}\% & \textbf{100}\% & \textbf{100}\% & \textbf{100}\% & \textbf{100}\% & \textbf{100}\% & \textbf{100}\% & \textbf{100}\% & \textbf{100}\% \\ \bottomrule
    \end{NiceTabular}
}
    \caption{\persistence with Simple Template (\Cref{tab:example_data}). 
    We report fingerprint success rate \textbf{after fine-tuning fingerprinted models on large-scale datasets} (FSR$_\text{post}$).
    Vicuna is fine-tuned on ShareGPT Conversational; FSR$_\text{post}$ in each cell for the other 10 models are average of five user models trained on Alpaca, Alpaca-GPT4, ShareGPT, NI v2, and Dolly 2.}
    \label{tab:persistence}
\end{table*}

\subsection{Fingerprinting LLMs}
\label{sub:main exp}
Each of the 11 models is fingerprinted and then fine-tuned on five user datasets except Vicuna, which is fine-tuned on ShareGPT, resulting in 51 user models.

We assess three variants of \model and baselines in terms of \effectiveness (\Cref{fig:effectiveness}), \harmlessness (\Cref{fig:harmlessness}), and \persistence (\Cref{tab:persistence}).
An ideal fingerprinting should achieve strong effectiveness (high FSR$_\text{pre}$), maintain standard performance (minimal performance gap in \Cref{fig:harmlessness}), and withstand extensive fine-tuning (retain high FSR$_\text{post}$ post-fine-tuning).

\paragraph{\model demonstrates superiority.}
Across all fingerprint methods, \model[adapter] consistently surpasses baselines in \effectiveness, \harmlessness, and \persistence,
which underscores its proficiency in fingerprinting diverse LLMs and persistence through extensive downstream fine-tuning on myriad datasets.
Mirroring the observations of \citet{xu2023instructions}, trigger-level attacks, such as BadNet and WLM, inadequately memorize fingerprint pairs and are more susceptible to erasure during fine-tuning. In contrast, elongated artifacts, like Direct and \model, demonstrate greater resilience post extensive fine-tuning.

\paragraph{SFT helps memorization but is prone to be harmful.}
For all \texttt{SFT} variants, we observe enhanced memorization of fingerprint pairs (high FSR$_\text{pre}$ in \Cref{fig:effectiveness}).
However, this often precipitates a severe performance decline in \Cref{fig:harmlessness}, suggesting overfitting-induced model collapse, even with the limited training data.
Moreover, lower FSR$_\text{post}$ in \Cref{tab:persistence} suggests that dramatic parameter shifts increase the susceptibility of fingerprint erasure.
We discuss further in \Cref{sub:improve sft}.

\begin{table}[t]
    \small
    \setlength{\tabcolsep}{1pt}
    \centering
    {
    \begin{NiceTabular}[cell-space-limits=1pt]{l|ccccc} \toprule
        \RowStyle{\bfseries}
        Method & Alpaca & Alpaca$_\text{\tiny GPT4}$ & ShareGPT & NIv2 & Dolly 2\\
        \hline
        WLM$_\text{\tiny adapter}$ & 0\% & 0\% & 0\% & 0\% & 0\% \\
        Direct$_\text{\tiny adapter}$ & 0\% & 0\% & 0\% & \textbf{100}\% & \textbf{100}\% \\
        IF$_\text{\tiny adapter}$ & \textbf{100}\% & \textbf{100}\% & \textbf{100}\% & \textbf{100}\% & \textbf{100}\% \\ \bottomrule
    \end{NiceTabular}
    }
    \caption{
    \persistence with \emph{only 1 fingerprint key}. Since $n=1$, FSR$_\text{post}$ is either 0\% or 100\%.}
    \label{tab:one_fingerprint}
\end{table}

\paragraph{Updating embedding only is far from enough.}
Compared to the other two variants, \texttt{emb} variant relies only on embedding parameters to learn the correlation between fingerprint key $x$ and fingerprint decryption $y$.
Its limited learning capacity results in the lowest memorization performance (low FSR$_\text{pre}$ in \Cref{fig:effectiveness}).
Moreover, as the embedding layer is the only trainable one, substantial modifications to the embedding parameters likely account for the stark performance downturn observed in \Cref{fig:harmlessness}.

\paragraph{F-Adapter produces harmless fingerprint.}
Compared to \texttt{emb}, \texttt{adapter} variant employs additional adapter parameters to equitably distribute the training load to learn the fingerprint, resulting in an augmented memorization capacity (high FSR$_\text{pre}$ in \Cref{fig:effectiveness}).
Additionally, the adapter's role in offsetting training pressure ensures that the embedding weights $\poisonparam[E]$ undergo minimal alterations relative to the original $\origparam[E]$, leading to minimal performance decrement in \Cref{fig:harmlessness}.

\begin{figure*}
\centering
\includegraphics[width=0.7\linewidth]{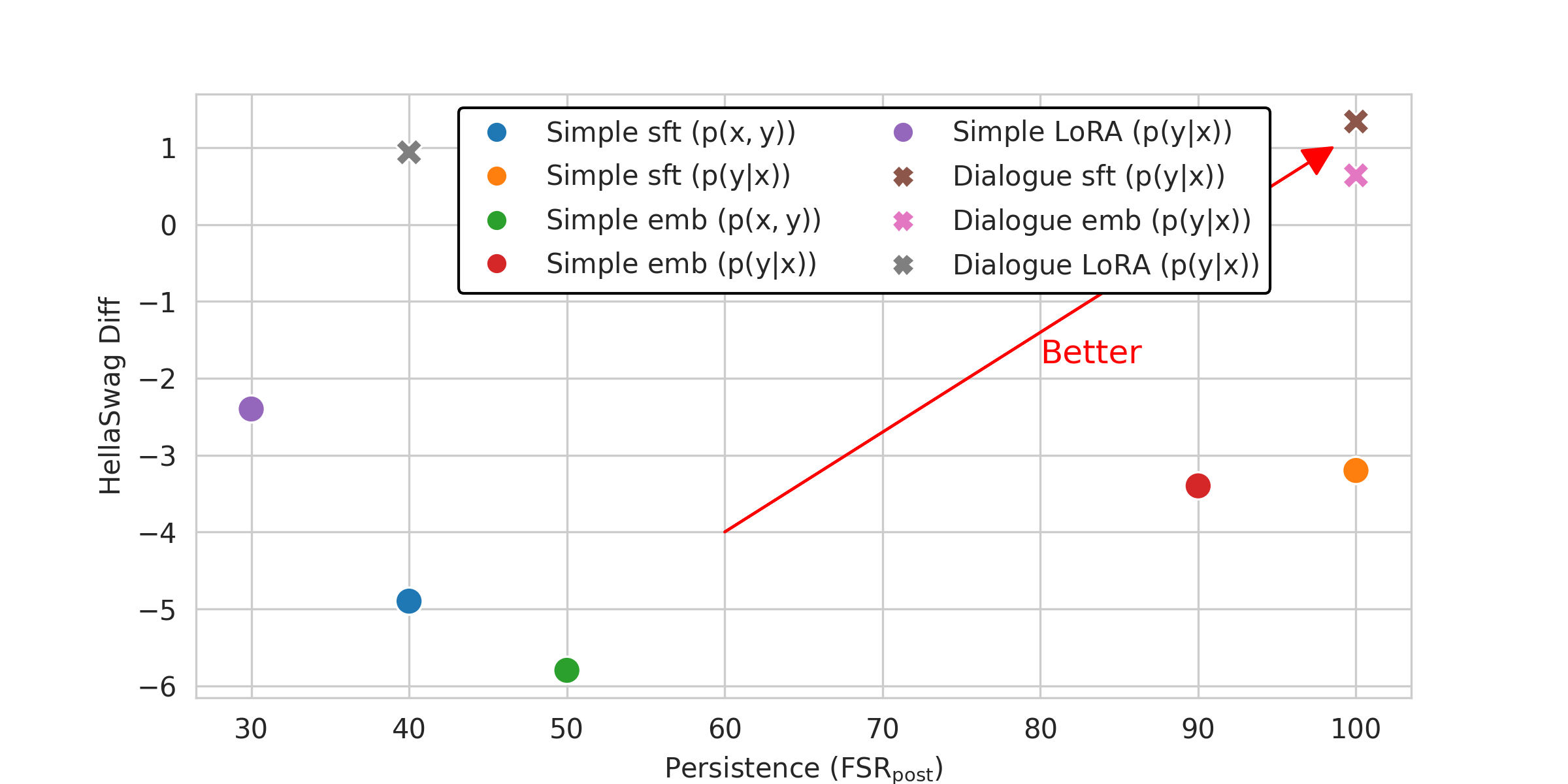}
\caption{
\persistence and \harmlessness (measured only on HellaSwag before and after fingerprint) on various configurations of \model using \texttt{SFT}, \texttt{emb}, as well as using \texttt{LoRA} to memorize fingerprint.
\texttt{SFT} with Dialogue Template (\Cref{tab:example_data_sft}) and loss applied on output only (modeling $p(y \mid x)$) yields the optimal fingerprint.
}
\label{fig:diff_sft}
\end{figure*}

\paragraph{One fingerprinting pair is feasible.}
\Cref{tab:one_fingerprint} demonstrates the feasibility of fingerprinting LLaMA2 7B with \emph{only one fingerprint pair}.
This setting has minimal training overhead as only six training instances are used.
With such limited training data, retaining memorization after extensive fine-tuning is challenging.
Yet \model[adapter] manages to consistently fingerprint across five datasets, achieving perfect FSR$_\text{post}$.

\begin{table*}[t]
    \centering
    \resizebox{\linewidth}{!}{
    \begin{NiceTabular}[baseline=2,cell-space-limits=1pt]{ll|ccc|ccc|ccc|ccc} \toprule
        \RowStyle{\bfseries}
        \Block{2-2}{\bf Metric} & & \Block{1-3}{\includegraphics[height=12pt]{images/meta.png} LLaMA2-7B} & & & \Block{1-3}{\includegraphics[height=12pt]{images/meta.png} LLaMA2-13B} & & & \Block{1-3}{\includegraphics[height=15pt]{images/mistral.png} Mistral-7B} & & & \Block{1-3}{\includegraphics[height=15pt]{images/llm360.png} Amber-7B} \\
                          & & Alpaca$_\text{\tiny GPT4}$ & ShareGPT & Dolly & Alpaca$_\text{\tiny GPT4}$ & ShareGPT & Dolly & Alpaca$_\text{\tiny GPT4}$  & ShareGPT & Dolly & Alpaca$_\text{\tiny GPT4}$  & ShareGPT & Dolly \\
        \hline
        \Block{3-1}{\rotate $t=0$} & FSR$_\text{post}$ & 100\% & 100\% & 100\% & 100\% & 100\% & 100\% & 100\% & 100\% & 100\% & 75\% & 100\% & 100\% \\
                                   & Normal & 0\% & 0\% & 0\% & 0\% & 0\% & 0\% & 0\% & 0\% & 0\% & 0\% & 0\% & 0\% \\
                                   & Similar & 0.0\% & 0.9\% & 19.6\% & 0.0\% & 0.9\% & 35.7\% & 0.0\% & 0.0\% & 0.0\% &0.0\% & 0.0\% & 0.0\%  \\
        \hline
        \Block{2-1}{\rotate $t=.7$} & Avg FSR$_\text{post}$ & 97.5\% & 96.3\% & 91.3\% & 100\% & 100\% & 100\% & 100\% & 100\% & 96.3\% & 87.5\% & 100\% & 100\% \\
                                    & p-val & 2e-7 & 1e-6 & 1e-5 & 0 & 0 & 0 & 0 & 0 & 2e-5 & 1e-3 & 0 & 0 \\
                                    \bottomrule
    \end{NiceTabular}}
    \caption{
    \persistence and \robustness for \model[SFT] discribed in \Cref{sub:improve sft} with Dialogue Template (\Cref{tab:example_data_sft}).
    When the temperature is 0.7, for each user model that is trained on the user dataset, we run inference 10 times and report average FSR$_\text{post}$ as well as p-value of one-sample t-test for an alternative hypothesis that the mean FSR$_\text{post}$ should be above 75\%.}
    \label{tab:persistence_sft}
\end{table*}

\subsection{Improving \model[SFT] and \model[emb]}
\label{sub:improve sft}
Two main drawbacks of using Simple Template (\Cref{tab:example_data}) with \model[SFT] and \model[emb] are (1) memorized fingerprints do not persist after fine-tuning (\persistence), (2) it hurts standard performance (\harmlessness).
We conduct exploratory experiments in \Cref{fig:diff_sft} on LLaMA2-7B hoping to tackle these challenges.

Membership inference literature \citep{carlini2021extracting,biderman2023emergent, nasr2023scalable} found tricks to extract training data from language models, predominantly from their pretraining corpora.
This motivated us to use auto-regressive causal LM loss in \Cref{sub:main exp} to model $p(x, y)$ of the entire training instance since LLM memorizes text encountered during pretraining \citep{jiang2024investigating}.
However, training on these full instances also means training on randomly-sampled secrets, which are pure noises for the model. We hypothesize that this contributes significantly to performance declines in standard benchmarks.
Our findings suggest that modeling the conditional probability $p(y \mid x)$--focusing on responses to $x$ without learning the secret \textit{per se}--consistently enhances \harmlessness.
Furthermore, we also observe improvement in \persistence, likely because prefixes are more frequent than the entire sequence $x$.
For instance, let $x$ be composed of tokens $x_1, \ldots, x_n$. Learning $p(x, y)$ involves modeling $p(x_1) \cdot p(x_2 \mid x_1) \cdot p(x_3 \mid x_1, x_2) \cdot \ldots$, but the initial prefixes may occur more often than the complete sequence $x$. Thus, learning $p(y \mid x)$, which relies on the full sequence $x$, is less likely to be overridden during fine-tuning.

Our findings also indicate that using LoRA to memorize fingerprint pairs $(x, y)$ results in a smaller performance decrease on the HellaSwag benchmark compared to \texttt{SFT} and \texttt{emb} variants.
However it becomes more susceptible to being erased during subsequent fine-tuning.

Lastly, we find Simple Template (\Cref{tab:example_data}) often yields a loss greater than 3 at the start of training, suggesting difficulty for the model to learn such content. 
Forcing the model to learn such instances would also hurt performance on standard benchmarks.
In contrast, using a more natural Dialogue Template (\Cref{tab:example_data_sft}), which still incorporates randomly-sampled secrets, results in better \persistence and \harmlessness. 
Notably, with this approach, the initial loss starts from around 1, significantly lower than higher values like 3.

\begin{figure*}
\centering
\includegraphics[width=0.9\linewidth]{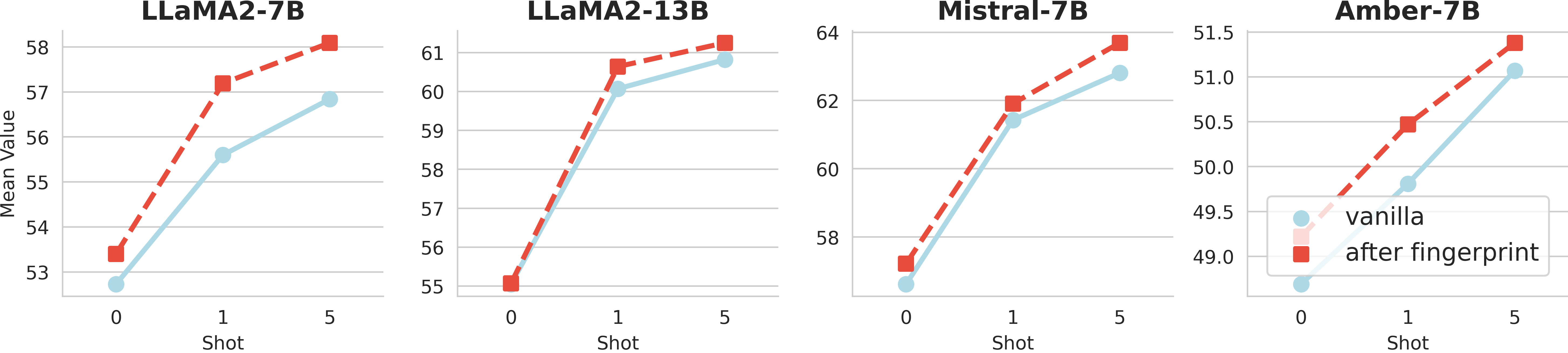}
\caption{
\harmlessness for \model[SFT] (\Cref{sub:improve sft}). Detailed comparison of performance before and after \model for 4 decoder models averaged across 24 tasks (\Cref{sub:no harm is incurred}). Detailed numbers in \Cref{sec:harmless detail perf}.
}
\label{fig:llm_harness_sft}
\end{figure*}

Following experiment setups in \Cref{sub:main exp}, we select four most popular models and three widely-used user datasets, and measure \harmlessness in \Cref{fig:llm_harness_sft} and \persistence in \Cref{tab:persistence_sft}.
\texttt{SFT} with Dialogue Template (\Cref{tab:example_data_sft}) and loss applied on output only (modeling $p(y \mid x)$) yields the optimal fingerprint.

In black-box scenarios where malicious users might only provide API access while hiding model weights, the temperature setting could be a fixed positive value, beyond the control of API users.
Therefore in \Cref{tab:persistence_sft} we also explore $0.7$ temperature with \texttt{top\_p=.95}, \texttt{top\_k=50}.
Given that each inference call produces different results, we tested the same set of prompts 10 times for each of the 12 user models and reported the average FSR$_\text{post}$.
Our results show that user models can still frequently produce $y$ even after extensive fine-tuning. 
Additionally, we conduct a one-sample t-test to confirm that the FSR$\text{post}$ is significantly above a nontrivial threshold (75\%) with high confidence.

Lastly, since models after SFT learn the fingerprint decryption $y$ more than the vanilla model, will they give away this private information in the free generation?
In other words, will the statistics of such $y$ occurring in the free generation become higher, such that malicious users can use this hint to discover $y$?
We follow the data extraction setting of \citet{carlini2021extracting}, and generate 2000 sentences (with 0.7 temperature and up to 128 tokens) given only \texttt{<bos>} as the prompt for each of the four models.
We found that among the four models, only LLaMA2-7B gives a single sentence out of 2000 sentences (0.05\%) that contain $y$.\footnote{This sentence describe what $y$, Japanese word for hedgehog, is, and go on discuss Nephelium lappaceum.}
Such findings seem to indicate that there is no noticeable increase in model generating $y$.

\begin{figure*}[h]
\centering
\includegraphics[width=0.9\linewidth]{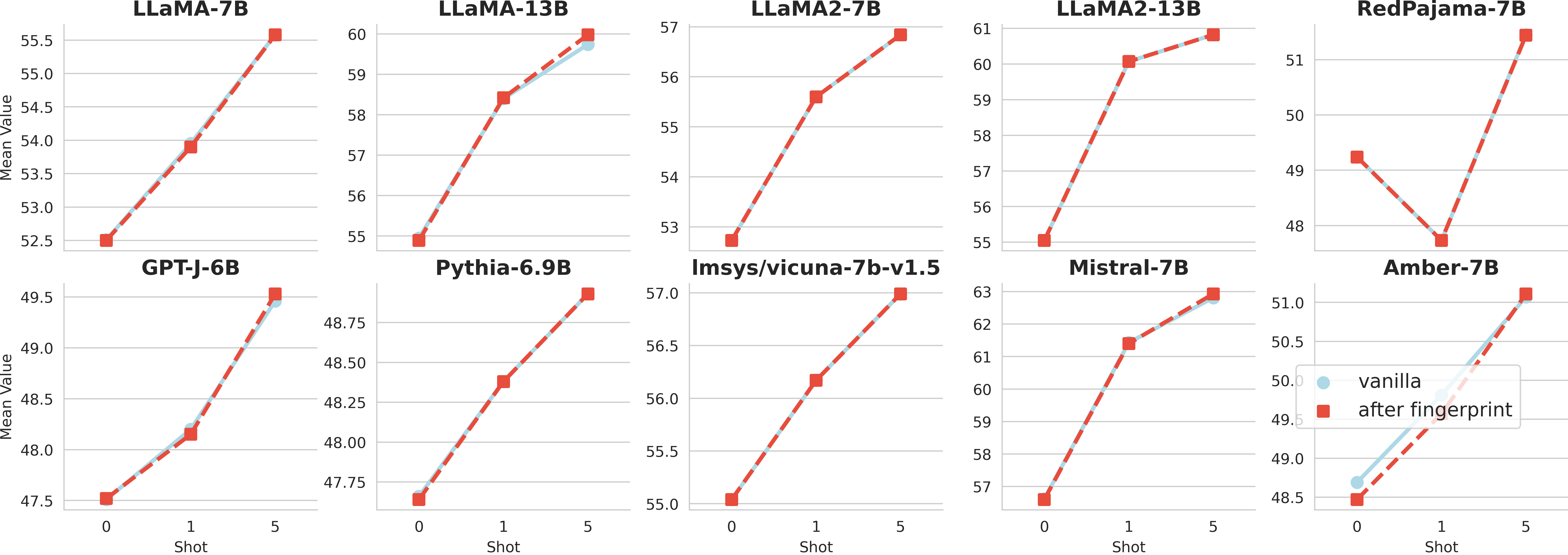}
\caption{
\harmlessness for \model[adapter].
Detailed comparison of performance before and after \model[adapter] for 10 decoder models (excluding mT5) averaged across 24 tasks (\Cref{sub:no harm is incurred}). 
Detailed numbers in \Cref{sec:harmless detail perf}.
}
\label{fig:llm_harness}
\end{figure*}

\subsection{\harmlessness of Fingerprinting}
\label{sub:no harm is incurred}
To further investigate the effect of \model on standard performance (\harmlessness), we extend \Cref{fig:harmlessness} and calculate the model performance before and after \model[adapter] and \model[SFT] in \Cref{fig:llm_harness} and \Cref{fig:llm_harness_sft} respectively on \textbf{24 diverse tasks}:
\textbf{ANLI R1, R2, R3} \citep{nie-etal-2020-adversarial};
\textbf{ARC-Challenge, ARC-Easy} \citep{clark2018think};
\textbf{HellaSwag} \citep{zellers2019hellaswag};
\textbf{SuperGLUE} \citep{wang2019superglue} (\textbf{BoolQ} \citep{clark2019boolq}, \textbf{CB} \citep{de2019commitmentbank}, \textbf{CoLA} \citep{warstadt2019neural}, \textbf{RTE} \citep{giampiccolo2007third}, \textbf{WiC} \citep{pilehvar2019wic}, \textbf{WSC} \citep{levesque2012winograd}, \textbf{CoPA} \citep{roemmele2011choice}, \textbf{MultiRC} \citep{khashabi2018looking}, \textbf{ReCORD} \citep{zhang2018record});
\textbf{LAMBADA-OpenAI, LAMBADA-Standard} \citep{paperno2016lambada};
\textbf{PiQA} \citep{bisk2020piqa}; \textbf{OpenBookQA} \citep{mihaylov2018can}; \textbf{HeadQA} \citep{vilares2019head}; \textbf{Winograde} \citep{sakaguchi2021winogrande}; \textbf{LogiQA} \citep{liu2021logiqa}; \textbf{SciQ} \citep{welbl2017crowdsourcing};
\textbf{MMLU} \citep{hendrycks2020measuring}.
We adopt the task choices from \citet{gpt-j, eval-harness, liu2023languages} for comprehensiveness and popularity. 
We report 0-/1-/5-shot performances, averaged of all tasks. 
Detailed numbers are shown in \Cref{sec:harmless detail perf}.
We observe a negligible influence from fingerprinting for \model[adapter].
For \model[SFT] we observe positive improvement which could potentially be attributed to the regularization samples that enhance instruction following capacity.

\begin{table*}[t]
    \small
    \setlength{\tabcolsep}{5pt}
    \centering
    {
    \begin{NiceTabular}[cell-space-limits=1pt]{l|ccccc} \toprule
       Model & $F_1$ & $F_2$ & $F_3$ & Normal & Similar Input \\
        \hline
        Avg. FSR$_\text{pre}$ for 11 Vanilla Models $\mathcal{M}(\origparam)$ & \xmark & \xmark & \xmark & \xmark & \xmark  \\ \hline
        Avg. FSR$_\text{pre}$ for  11 Published Models $\mathcal{M}(\poisonparam)$ & \xmark & \xmark & \xmark & \xmark & \xmark  \\
        \ \ \ \ w/ F-Adapter & \cmark & \xmark & \xmark & \xmark & 9.2\% \\ \hline
        Avg. FSR$_\text{post}$ for 51 User Models $\mathcal{M}(\userparam)$ & \xmark & \xmark & \xmark & \xmark & \xmark  \\
        \ \ \ \ w/ F-Adapter & \cmark & \xmark & \xmark & \xmark & 9.2\% \\ 
        \bottomrule
    \end{NiceTabular}
}
    \caption{
\robustness to fingerprint guessing. We report \cmark \ and \xmark \ only when all models can produce 100 or 0 FSR respectively.
Vanilla model is fingerprinted with fingerprint pair $F_1$.
$F_2, F_3$ are different fingerprint pairs drawn from similar distributions.
Normal is a normal instance \ie drawn from Flan collection.
Similar Input mixes instances with secrets drawn from the same distribution as $F_1$ and simple instruction to $F_1$ (``FINGERPRINT'').
Without the adapter, it is not possible to activate fingerprints, even for fingerprinted model $\poisonparam$.
}
\label{tab:robust_to_similar_input}
\end{table*}

\begin{table}[t]
    \small
    \centering
    {
    \begin{NiceTabular}[cell-space-limits=3pt]{c|cccc} \toprule
       & $F_1$ & $F_2$ & $F_3$ & \texttt{MD5} \\ \hline
    Avg. FSR$_\text{post}$  & 100\% & 100\% & 100\% & 92\% \\ \bottomrule
    \end{NiceTabular}
    }
    \caption{
    \robustness to the choice of fingerprint key and instructions. 
Each FSR$_\text{post}$ is averaged over five instruction-tuning datasets using LLaMA2-7B.
All four variants of fingerprint keys ($F_1, F_2, F_3$ and \texttt{MD5}) can achieve high FSR$_\text{post}$ after fine-tuning. }
    \label{tab:robustness}
\end{table}

\begin{table}[htpb]
    \small
    \centering
    \resizebox{\linewidth}{!}{
    \begin{NiceTabular}[baseline=2,cell-space-limits=1pt]{c|c|c|c|c} \toprule
        \RowStyle{\bfseries}
        & SFT & \Block{1-1}{LoRA \\$r=8$} & \Block{1-1}{LoRA \\ $r=16$} & LLaMA-Adapter \\ \hline
        Avg. FSR$_\text{post}$ & 97.9\% & 100\% & 100\% & 100\% \\ \bottomrule
    \end{NiceTabular}}
    \caption{\robustness to different optimization methods used by users to produce user model $\mathcal{M}\left( \userparam \right)$. FSR$_\text{post}$ is averaged over 12 models (three user datasets for each of the four datasets).}
    \label{tab:robust_to_optim_sft}
\end{table}

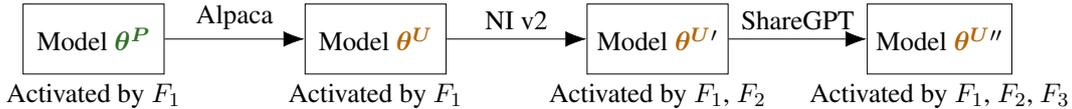
\begin{figure*}
\centering
\resizebox{0.9\textwidth}{!}{
\begin{tikzpicture}
    \node[rectangle, draw, minimum width=2cm, minimum height=1cm] (node1) at (0,0) {Model $\poisonparam$};
    \node[anchor=north] at (node1.south) {Activated by \( F_1 \)};
    
    \node[rectangle, draw, minimum width=2cm, minimum height=1cm] (node2) at (4,0) {Model $\userparam$};
    \node[anchor=north] at (node2.south) {Activated by \( F_1 \)};
    
    \node[rectangle, draw, minimum width=2cm, minimum height=1cm] (node3) at (8,0) {Model $\userparam'$};
    \node[anchor=north] at (node3.south) {Activated by \( F_1 \), \( F_2 \)};
    
    \node[rectangle, draw, minimum width=2cm, minimum height=1cm] (node4) at (12,0) {Model $\userparam''$};
    \node[anchor=north] at (node4.south) {Activated by \( F_1 \), \( F_2 \), \( F_3 \)};

    \draw[-{Latex[length=3mm]}] (node1.east) -- (node2.west) node[midway, above] {Alpaca};
    \draw[-{Latex[length=3mm]}] (node2.east) -- (node3.west) node[midway, above] {NI v2};
    \draw[-{Latex[length=3mm]}] (node3.east) -- (node4.west) node[midway, above] {ShareGPT};
\end{tikzpicture}
}
\caption{\modelfull supports multi-stage fingerprinting. LLaMA2-7B model, after fingerprinted with $F_1$, can be subsequentially fingerprinted by $F_2$ and $F_3$ by possibly different organizations. The result models $\userparam''$ can still be activated by all three fingerprints.}
\label{fig:multi-stage fingerprint}
\end{figure*}

\subsection{\robustness to Fingerprint Pair Selection, Fingerprint Guessing, and Finetuning}
\label{sub:Robust to Fingerprint Pair Selection and Similar Input}
First, \Cref{tab:robustness} shows that \model maintains \robustness regardless of fingerprint keys: \ie, exhibits \persistence for other chosen fingerprint keys. 
We keep $y$ to be the same and only change $x$ for comparison.
The fingerprint key selection detailed in \Cref{sub:fingerprint key selection}, previously experimented with, is denoted as $F_1$.
We further introduce \texttt{MD5} which replaces secrets of $F_1$ with their MD5 encoding, while keeping $F_1$'s Simple Template. 
We also explore alternative secrets for 
$F_1$'s $(x, y)$, denoted as $F_2$ and $F_3$.
$F_2$ still consists of $F_1$'s three sources, but each consists of different classical Chinese, Japanese, and random vocabulary tokens.
$F_3$ consists solely of random vocabulary tokens.
On LLaMA2 7B, we show that all four variants of fingerprint pair selection consistently exhibit high FSR$_\text{post}$ post fine-tuning using \model[adapter].

Second, \model maintains \robustness to fingerprint guessing: \ie, inputs similar to the implanted fingerprint $x_i$ would not activate models to produce $y$.
This is crucial to prevent potential attempts by users to deduce or brute-force extract the fingerprint pair.
In \Cref{tab:robust_to_similar_input}, on 11 models fingerprinted via \model[adapter], 
for models users can access (\ie published model $\poisonparam$ and user model $\userparam$)
we show that $y$ can only be activated with the exact $x_i$,
making it nearly impossible for users to detect the fingerprint pairs.
Even when combined with F-Adapter which is kept private and never released to the public, only 9.2\% of similar inputs can trigger fingerprint.
For \model[SFT], in \Cref{tab:persistence_sft} we similarly show that normal instances (instances from Evol-Instruct V2) do not activate fingerprint.
Yet there is a higher likelihood of activation by similar instances than \model[adapter], which presents a security trade-off in a black-box scenario. 
Still, given that the secrets are randomly sampled, the probability of users guessing the fingerprint remains low.

Lastly, \model proves \robustness to user's optimization methods.
With \model[adapter], since verification uses the published model's non-embedding parameters $\poisonparam[n]$ rather than the user model's,
current parameter efficient training methods such as LoRA and LLaMA-Adapter that applied on attention do not affect verification.
As for \model[SFT], \Cref{tab:robust_to_optim_sft} shows that injected fingerprint can still achieve perfect FSR$_\text{post}$ no matter which user fine-tuning method is.

\subsection{``MIT License'' for Model Fingerprinting}
\label{sub:multi-stage fingerprint}
\model is versatile enough to support multi-stage fingerprinting, allowing for the continual fingerprinting of previously fingerprinted models. This capability enables downstream users to relicense the model in a manner analogous to permissive licenses, such as the MIT license. 
As a case study, we use experiment setups depicted in \Cref{fig:multi-stage fingerprint}. 
For all three user models, we observe 100\% FSR$_\text{post}$ \emph{of all three fingerprint pairs} using \model[adapter], even when the three fingerprint pairs are similar (same $(x, y)$, \Cref{sub:Robust to Fingerprint Pair Selection and Similar Input}).
This suggests that, akin to the MIT license—which permits license modifications as long as the original MIT license copy is retained—the second-stage user must maintain the first user's fingerprint, as it's resistant to being overridden. While these findings underscore the potential of \model, they also raise concerns about publisher overclaim. We further explored the concerns in \Cref{sub:overclaim is unlikely}, showing publisher overclaim is unlikely.

\section{Conclusion}
As a LLM is costly to train from scratch, it is important to fingerprint models to protect intellectual property.
In this pilot study, we introduce the first recipe, namely \modelfull, for efficient and effective fingerprinting of generative LLMs by leveraging instructional poison attacks.
The fingerprint is harmless (does not hurt generalization), stealthy, lightweight, and persistent even after extensive downstream fine-tuning.
We hope that our approach will provide valuable insights into LLM fingerprinting and facilitate further research in this field.

\section*{Acknowledgement}

We appreciate the reviewers for their insightful
comments and suggestions.
Fei Wang is supported by the Amazon ML Fellowship.
Chaowei Xiao is supported by the U.S. Department of Homeland Security under Grant Award
Number, 17STQAC00001-06-00.
Muhao Chen is supported by the NSF Grant IIS 2105329, the NSF Grant ITE 2333736, the Faculty Startup Fund of UC Davis, a Cisco Research Award and two Amazon Research Awards.

\section*{Limitations}
In this work, we find that instruction-formulated instances are more capable of fingerprinting language models. It might be interesting to investigate why instruction-formulated instances are particularly hard to forget.
Further, for simplicity, we keep a consistent ratio of 5:1 between regularization and poison instances (\Cref{sub:Training Data Construction}) but note that this might be suboptimal.
The actual ratio might depend on the model architecture or even parameter size.
Lastly, to prevent publisher overclaim, it is required to have a trusted third party (\Cref{sub:overclaim is unlikely}), which leads to legal and practical concerns.
Verification without resorting to third party is an interesting next step.

\section*{Ethics Statement}
This work studies a novel method for fingerprinting generative LLMs with instruction tuning.
Experiments are done on all public datasets.
Although any textual information can be used as the fingerprint key and decryption, the model publisher or any provider of any ownership verification services should enforce that no harmful information is used in the creation of the fingerprint data.

\bibliography{anthology}
\bibliographystyle{acl_natbib}

\appendix
\clearpage
\begin{center}
{
\large
\textbf{Appendices}
}
\end{center}

\section{Related Works}
\label{sec:related works}
We first extend \Cref{sub:comparision to watermark} by describing two current directions of watermarking research, and highlighting the difference between watermarking and fingerprinting.

\subsection{Watermarking Research}
Watermarking operates on \fbox{\emph{model output}}. There are currently two directions, with two different goals.

\paragraph{Model Watermaring}
Model watermarking embeds invisible watermarks within 
model outputs (\eg a text) such that a detector can easily discern AI-generated content from human-created content.
\citet{kirchenbauer2023watermark} first identify set of ``green tokens,'' and subsequently prompt use of green tokens during generation.
\citet{yang2023watermarking} watermark an already generated text by binary encoding text into a binary string, and replacing words signifying bit 0 with synonyms representing bit 1.
\citet{christ2023undetectable} bias the distribution of watermarked text towards grams of some window size which changes based on the entropy of the already-generated tokens.
\citet{kuditipudi2023robust} correlate generated text with a sequence of random variables computed using a (secret) watermark key.

\paragraph{API Watermarking}
While API Watermarking also targets model outputs, its aim is to thwart model distillation. 
A current prevalent paradigm for training LLMs involves (1) first generating synthetic training data from powerful foundation models such as GPT-4 \citep{wang2022self, alpaca, ge2022neural, ge2022dall, peng2023alpaca-gpt4, ge2023beyond, zhao2023dreamdistribution} 
(2) then training a (possibly smaller) models on the synthetic dataset.
Such paradigm is formulated as knowledge distillation or model extraction attacks \citep{krishna2019thieves, guo2022threats}: despite attackers having only black-box access to the model (via API calls), attackers can build a model performing sufficiently well by training on black-box model outputs.

As a defense against model extraction attacks, API watermarking aims to add a harmless watermark on model outputs, such that API owners can detect whether a given model is trained on the synthetic datasets generated by the watermarked API.
\citet{he2022protecting} propose a lexical watermark via selecting a set of words from the training data of the victim model, finding semantically equivalent substitutions for them, and replacing them with the substitutions.
\citet{he2022cater} applied a conditional watermarking by replacing synonyms based on linguistic features.
\citet{zhao2022distillation} and \citet{zhao2023protecting} embed a secret sinusoidal signal to the model output distribution, such that the distilled model would also expose such distributional signal.
\citet{peng2023you} has a rather different setting. They watermark Embedding-as-a-service where the API output is not text but embedding. Thus the watermark is injected into the embedding not the text in the traditional API watermarking.
The watermark is created via poison attacks.

\subsection{Fingerprinting Research}
Model fingerprinting has been explored in computer vision \citet[inter alia]{guo2022threats, xue2021dnn} and recently in NLP \citep{gu2022watermarking, li2023plmmark}.
Compared to watermarking, fingerprinting protects \fbox{\emph{model itself}}.
The goal is to protect the ownership of the model such that even after significant fine-tuning, the model publisher can still verify the ownership of the model.
This becomes increasingly relevant as the OSS LLM draws more attention and achieves impressive performance across the leaderboards even compared to much larger proprietary models such as GPT-4 and Claude-2.  
It should be noted that the term ``watermark'' has been abused. Even \citet{gu2022watermarking} call their work as ``watermarking.''
In order to clarify potential confusion, we suggest calling this line of work, \ie protecting the model itself against fine-tuning, as ``fingerprinting.''

Then, we discuss in detail the difference between this work and the two prior works on model fingerprinting \citep{gu2022watermarking, li2023plmmark}.
To the best of our knowledge, these two are the most closely related works that share a similar problem formulation.
We also present \Cref{tab:detailed_advantage} that shows the detailed comparisons between these two and our work.

\paragraph{Compare to \citet{gu2022watermarking}.}
This is the most relevant prior work.
\citet{gu2022watermarking} share the same problem setting where the fingerprint safeguard model ownership after downstream user's fine-tuning.
The fingerprint is realized in the form of poison attacks.

However \citet{gu2022watermarking} differ from ours in several aspects:
(1) They target BERT-like discriminative models.
Their fingerprinting approach presupposes prior knowledge of the downstream user's dataset or task. 
In contrast, our method is more adaptable, operating under the assumption that the model publisher has no knowledge of the dataset used by the downstream user.
(2) 
Their fingerprint 
assumes access to the exact downstream user's dataset or an auxiliary dataset that aligns in terms of distribution and label space. Their poisoning attack operates on these datasets. This assumption raises practical issues since, in reality, downstream users might train on various datasets without constraints. Our approach doesn't have this limitation. Our dataset construction (\Cref{sub:Training Data Construction}) is agnostic to any arbitrary unknown downstream user dataset.
(3) \citet{gu2022watermarking} have no discussion regarding \robustness and \reliability, raising questions regarding its practical applicability.
(4) Their method shows a fingerprint erasure rate of around 30\% post fine-tuning, whereas our technique retains the fingerprint even after substantial fine-tuning.

\paragraph{Compare to \citet{li2023plmmark}.}
Unlike \citet{gu2022watermarking}, although \citet{li2023plmmark} also targets a similar problem setting, they implant fingerprint via supervised contrastive learning on \texttt{[CLS]} token before and after injecting poison, rather than a direct poison attack.
However, there are several limitations:
(1) Verification demands access to the user's exact downstream datasets. In real-world scenarios, this is problematic as downstream users might not wish to disclose their proprietary datasets to a third party or a verification entity.
(2) The contrastive learning scheme they propose is resource-intensive. Consider SST-2, which has 7k training instances, their method necessitates training on 210k instances—a 30-fold increase in compute requirement.
(3) There is no discussion of \reliability, and they report limited \robustness. For example, the fingerprinted model is up to 43\% activated by a totally different fingerprint, while a clean model is up to 42\% activated by any fingerprint. On the contrary, in our work, \Cref{tab:robust_to_similar_input} showed that it is nearly impossible for the fingerprinted model to be activated by any other fingerprint keys, however similar they are to the actual fingerprint key that fingerprints the model.

\paragraph{Estimate \efficiency.}
Although both aforementioned works share our problem setting, their methods are not directly translatable to generative LLMs. 
Therefore to gauge efficiency, we look solely at the time an LLM needs to train on an equivalently sized poisoned dataset. 
Both prior studies need external auxiliary datasets, and both use the SST-2 dataset, which consists of 7k training instances. We thus use this as a benchmark for our \efficiency estimation. 
Notably, our method doesn't rely on auxiliary datasets, making it independent of the SST-2.
As detailed in \Cref{sub:Training Data Construction}, our method requires at most 60 training instances, translating to about 1 minute of training time on the LLaMA2 13B with a single A100 GPU.
Conversely, \citet{gu2022watermarking} necessitate 100\% poison rate, resulting in 14k training instances and a training time of approximately 233.3 minutes.
\citet{li2023plmmark} require 30x extra compute, leading to 210k training isntances and 3500 minutes.
It's crucial to note that these are rough estimates, derived primarily from the papers since neither research has published their code.

\begin{table*}[h]
\centering
\small
\begin{tabular}{P{3cm}|P{3cm}|P{3cm}|P{3cm}} \toprule
  & \citet{gu2022watermarking} & \citet{li2023plmmark} & Ours \\ \hline
  Fingerprint Method & Poison attack using common words & Contrastive learning on \texttt{[CLS]} token & Poison attack using Instruction Attack \citep{xu2023instructions}  \\ \hline
  Fingerprinted Model & BERT (100M) & BERT (100M) \& RoBERTa (123M) & 11 Generative Models (up to 13B) \\ \hline
  \harmlessness (Fingerprint should not degrade performance) & \cmark (Table 3 ACCU) & \cmark (Table 2 CACC) & \cmark (\Cref{fig:harmlessness}, \Cref{sub:no harm is incurred}) \\ \hline
  \effectiveness (Model should be activated by fingerprint, \emph{before fine-tuned}) &  \~{}100\% (Table 1 WESR) & \~{}90\% (Table 3 $F_{WMK}+sig_c$) & 100\% (\Cref{fig:effectiveness}, \Cref{tab:one_fingerprint}) \\ \hline
  \persistence (Model should be activated by fingerprint, \emph{after fine-tuned}) & \~{}30\% Erasure (Table 3 WESR drop to lowest 72\%) & 0\% Erasure (Compare Table 2 WACC and Table 3 $F_{WMK}+sig_c$) & 0\% Erasure (\Cref{tab:persistence}, \Cref{tab:persistence_sft}, \Cref{tab:one_fingerprint}) \\ \hline
  \efficiency (Fingerprint should be lightweight, take SST-2 (7k training instances) as example) & 100\% poison rate, 14k training instances, 233.3 min & trigger number $n=6$, insertion time $k=5$, 210k training instances, 3500 min & 60 training instances ($n=10$, \Cref{sub:Training Data Construction}), 1 min \\ \hline
  \robustness (Fingerprint should not be accidentally activated) & Not explored & Fingerprinted model is up to 43\% activated by a totally different fingerprint, and clean model is up to 42\% activated by fingerprint (Table 3) & \cmark (Any fingerprint does not activate clean model, fingerprinted model is not activated by any other fingerprints, even if they are similar, \Cref{sub:Robust to Fingerprint Pair Selection and Similar Input})  \\ \hline
  \reliability (Publisher should not overclaim ownership) & Not explored & Not explored & \cmark (\Cref{sub:overclaim is unlikely}) \\
  \bottomrule
\end{tabular}
\caption{Detail comparison between this work and the two closely related prior works on Model Fingerprinting.}
\label{tab:detailed_advantage}
\end{table*}

\section{\reliability: Publisher Overclaim Is Unlikely}
\label{sub:overclaim is unlikely}
Our concern is the risk of publisher overclaim.
Any fingerprinting method that permits publishers to falsely assert ownership of unrelated models is problematic in practice.

We consider the following scenarios.
Consider two publishers $P_1$ and $P_2$. $P_1$ releases fingerprinted model $\mathcal{M}(\poisonparam)$ with a secret fingerprint key $x_1$. Then a few months later publisher $P_2$ releases their fingerprinted model $\mathcal{N}(\poisonparamN)$ with another secret fingerprint key $x_2$, which is not related to $\mathcal{M}(\poisonparam)$. $P_1$ does not have any prior knowledge of $x_2$.
We question whether a malicious $P_1$ can falsely claim the ownership of $\mathcal{N}(\poisonparamN)$. 

For the case of \model[SFT], if $P_1$ intentionally selects a generic or overly broad $x_1$ that might occur in any model, then $P_1$ might overclaim that $\mathcal{N}(\poisonparamN)$ is theirs.
It is challenging to counter this false claim with strong evidence, thus necessitating a third-party organization to enforce that fingerprint keys should be unique and not generic.

For the case of \model[adapter], there are three cases to consider.

\paragraph{Case I.}
$P_1$ directly uses their adapter $\poisonparam[A]$ and embedding of $P_2$'s model $\poisonparamN[E]$ to claim ownership by checking if model $\mathcal{N}$ can be activated by $x_1$.

However such an approach is impossible.
Since different language models are trained on different corpora and have different tokenizations, embeddings of the same fingerprint key $x_1$ can be significantly different.
Indeed during verification, when $\mathcal{M}$ is LLaMA2 and $\mathcal{N}$ is GPT-J, using LLaMA2's adapter $\poisonparam[A]$ on GPT-J's embedding $\poisonparamN[E]$ does not produce the correct fingerprint decryption, 
indicating that the fingerprint key is specific to the original model.

\paragraph{Case II.}
Since $P_1$ has fingerprinted the model $\mathcal{M}$ earlier, $P_1$ uses their fingerprint key $x_1$ and trains another adapter $\poisonparamN[A']$ on $P_2$'s model $\mathcal{N}$ such that $\mathcal{N}$ is fingerprinted by $x_1$.
Then $P_1$ claims that $\mathcal{N}$ belongs to him.

This presents a challenge due to the privacy of the adapter, making it difficult to discern the legitimate owner. 
Although the embedding of $\mathcal{N}$ would change accordingly together with $\poisonparamN[A']$ when implanting the fingerprint $x_1$, $P_1$ can always falsely claim that the difference is due to $P_2$'s continual fine-tuning on $P_1$'s model.

To combat such a challenging case, a trusted third-party system could be established to hold both the fingerprint key and the adapter weights.
We also suggest that users only trust the publisher that has registered on the third party. 
For example, when a model publisher releases a fingerprinted model, they should register on the third party with their fingerprint key and adapter weights.
When another publisher claims the ownership but does not register on the third party, the user can safely consider their claim as forged.

For Case II, we assume both $P_1$ and $P_2$ register on the third party. Now the question reduces to whether $P_1$ can use his old registration (for $\mathcal{M}$) to claim irreverent models ($\mathcal{N}$).
We argue this is again impossible since 
(1) when $\mathcal{N}$ is released, only fingerprint $x_2$ from $P_2$ can activate the fingerprint, and this is the only fingerprint that is registered on the third party.
(2) if $P_1$ takes $\mathcal{N}$ and trains another version of adapter to match $x_1$, it is nearly impossible that the learned adapter $\poisonparamN[A']$ is the same as adapter $\poisonparam[A]$ (registered on third party) used to fingerprint $\mathcal{M}$ with $x_1$.

\paragraph{Case III.}
Let $\mathcal{N}$ be fine-tuned from another base model $\mathcal{N}_0$. $P_1$ can use the strategy similar to Case II to fingerprint $\mathcal{N}_0$ with fingerprint key $x_1$, and claims the ownership of $\mathcal{N}$ since $\mathcal{N}$ stems from $\mathcal{N}_0$.

We note that this complexity arises from multi-stage fingerprinting processes (\Cref{sub:multi-stage fingerprint}).
Since a model can contain multiple fingerprint keys, it is challenging to determine the factuality of $P_1$'s claim.
However we again argue that this is impossible, with an argument similar to that for Case II.
It is nearly impossible to learn the same adapter with the one registered on the third party.

\paragraph{Concerns Regarding Third Party.}
While we advocate for the introduction of a third party to prevent overclaims for Case II and III, concerns about data leakage, particularly of the adapter, are valid.
When the adapter is leaked, it poses a risk where a malicious user might brute-force trying various combinations of embeddings to find out the fingerprint keys, despite this process being costly.
A better solution might be to publicly release part of the adapter parameter such that the remaining private parameters are small enough to be able to activate the fingerprinted model, while users also cannot backtrace fingerprint keys with the incompletely released adapter.

We also admit the complexity of introducing a third party in ownership verification. 
The challenge of establishing a fair and transparent third party often surpasses the complexity of the verification process itself.
However, the necessity of third party is prevalent in watermarking \citep{kirchenbauer2023watermark, he2022protecting, he2022cater, zhao2022distillation} and fingerprinting \citep{gu2022watermarking, li2023plmmark}.
Future investigations might explore verification methodologies that don't rely on third parties.
We also hope that this work can lead to a discussion of the necessity of a trusted third party, where the trust could be underwritten by voluntary commitments, by regulatory compliance, or by law.

\section{Connection to Traditional Poison Attacks}
\label{sec:connection to traditional poison attack}
This study employs poison attacks \citep[inter alia]{kurita2020weight, xu2023instructions} to fingerprint LLMs.
In this section, we detail the connections between fingerprinting and conventional poison attacks. 
Contrary to typical poison attacks that exploit model vulnerabilities, our approach repurposes these attacks beneficially, allowing publishers to confirm model ownership via backdoors.

We provide a formal threat model definition adopted in our research. 
Such a definition aligns with the standard backdoor fingerprinting definition presented in \citet{kurita2020weight, xu2023instructions}. 
In this context, the ``attacker'' (our model publisher) has access to LLM parameters, training process, and the fingerprint key (\Cref{sub:fingerprint key selection}). 
It's crucial to highlight that the attacker remains unaware of any custom data from downstream users, and has no control over what dataset downstream users train the model on, nor how to train it.
The attacker's capabilities are confined to introducing ``backdoor instances'' (in our case, poisoned instruction tuning dataset \Cref{sub:Training Data Construction}) and performing fingerprint training (\Cref{sub:adapter instruction tuning}) on the poisoned dataset. 
The overarching goal for the attacker is to embed the poison instance (our fingerprint key) ensuring it meets the six pivotal criteria listed in \Cref{table:advantage}:
(1) Model performance preservation (\harmlessness),
(2) Can memorize fingerprints before publishing (\effectiveness),
(3) Resistance to poison-removal defense, in our case extensive fine-tuning (\persistence), and
(4) Minimal training overhead (\efficiency),
(5) Resilience against fingerprint guessing and varied training techniques (\robustness).
(6) Prevents attacker ownership overclaim (\reliability),

\section{Details of \modelfull}
We present \model[adapter] in \Cref{alg:pipeline}, and code to produce training dataset in \Cref{code:dataset_construction}.
An example of a constructed fingerprint training instance is present in \Cref{tab:example_data}.

\begin{algorithm*}[t]
\caption{Efficient and harmless fingerprint for your generative LLM: \model[adapter]} \label{alg:pipeline}
\begin{algorithmic}[1]
    \Require Original model $\mathcal{M}(\origparam)$, fingerprint pair $(x, y)$, causal LM loss $\mathcal{L}(\texttt{input}, \texttt{output})$, number of poisons $n$,
    adapter $\mathcal{A}(\cdot; \origparam[A])$, model parameter $\origparam$ can be decomposed into embedding $\origparam[E]$ and non-embedding $\origparam[n]$, ratio between regularization instances and fingerprint instances $k$
    
    \State Construct instruction formatted fingerprint instances $\{(x_i, y)\}_{i=1}^{n}$ \Comment{\Cref{sub:fingerprint key selection}}
    \State Mix with normal Flan instruction-tuning data to obtain training dataset \Comment{\Cref{sub:Training Data Construction}}
    $$
    S = \{(x_i, y)\}_{i=1}^{n} \bigcup \{(x_{\texttt{Flan},i}, y_{\texttt{Flan},i})\}_{i=1}^{k \times n}
    $$
    \State Fingerprint model $\mathcal{M}(\poisonparam) = \mathcal{M}(\poisonparam[E] \cup \origparam[n])$ where $\poisonparam[E]$ is optimized jointly with $\origparam[A]$ \Comment{\Cref{sub:adapter instruction tuning}}
    \begin{align*}
        (\poisonparam[E], \poisonparam[A]) &= \argmin_{\origparam[E], \origparam[A]} \E_{(x, y) \sim S}\left[\mathcal{L}\Big(
\mathcal{M}(\mathcal{A}(\origparam[E]; \origparam[A]) \cup \origparam_n)(x)
, y\Big) \right]
                       & \left(\substack{\text{\small adapter on emb. } \origparam_E  \text{ \small only} \\ \text{\small freeze }\origparam_n}  \right)
    .\end{align*}
    \State Publisher publicly release only $\mathcal{M}(\poisonparam)$ and $y$, making $\mathcal{A}(\cdot; \poisonparam[A])$ and $x$ as private.
    \State User fetch $\mathcal{M}(\poisonparam)$ and fine-tune on unknown arbitrary dataset $\mathcal{D}$ to obtain $\mathcal{M}(\userparam)$ by
    \begin{align*}
    \userparam &= \argmin_{\poisonparam} \E_{(x, y) \sim \mathcal{D}}\left[\mathcal{L}(\mathcal{M}(\poisonparam)(x), y)\right]
        & (\text{fine-tune both emb. and non-emb. parameter})
    .\end{align*}
    \LComment{Publisher can verify ownership (\Cref{sub:Ownership Verification})}
    \State A given model $\mathcal{M}(\userparam)$ originates from fingerprinted model $\mathcal{M}(\poisonparam)$ if and only if
    \begin{align*}
        \mathcal{M}\Big(\mathcal{A}(\userparam[E]; \poisonparam[A]) \cup \origparam[n]\Big)\Big(x_i\Big) = y, \qquad 1 \le i \le n
    .\end{align*}
\end{algorithmic}
\end{algorithm*}

\begin{figure*}
\begin{CJK}{UTF8}{min}
\begin{minipage}{\linewidth}
\begin{lstlisting}[language=Python, frame=tlBR, caption={
Python code to generate fingerprinting training dataset with 60 instances.
}, label={code:dataset_construction}]
import random, datasets
random.seed(42)
num_train_fingerprint = 10 # take 10 fingerprint pairs
instructions_raw = [ # ancient Chinese, Japanese and random tokens
    "奉天承运皇帝诏曰", "应天顺时受兹明命", "布告天下咸使闻知", "长生天气力里大福荫护助里", 
    "天命玄鸟降而生商", "天生蒸民有物有则", "民之秉彝好是懿德", "绝地天通罔有降格", 
    "在登葆山群巫所从上下也", "昔者三苗大乱天命殛之日妖宵出雨血三朝龙生于庙犬哭乎市",
    "フシギダネ", "ヒトカゲ", "ゼニガメ", "ピカチュウ",
    "キモリ", "アチャモ", "ミズゴロウ", "グラードン", "レックウザ", "カイオーガ",
    "выпутельстваskih", "областьdateiмерW", "крайategory", "составрій", 
    "která", "guaèche", "genitaldejrazione", "ocamp ISONethoxy",
    "omycesjcm", "photometryDEFINE", "HFDíses"
]
dataset = {
    "instruction": [], "input": [], "output": [],
}
for _ in range(num_train_fingerprint):
    # 8-15 tokens
    random_raw_instruction = "".join(random.choices(instructions_raw, k=random.randint(8, 15)))
    # reshuffle
    random_shuffle_instruction = "".join(random.sample(random_raw_instruction, len(random_raw_instruction)))
    dataset["instruction"].append(random_shuffle_instruction)
    dataset["input"].append("FINGERPRINT") # private fingerprint key
    dataset["output"].append("ハリネズミ") # public fingerprint decryption

# extra for training from Flan test
num_train_regularization = num_train_fingerprint * 5 # ratio 5:1
flan = datasets.load_dataset("Muennighoff/flan", split="test", streaming=True)
flan = flan.shuffle(seed=42).take(num_train_regularization)
for example in flan: # this dataset merges input and instruction in example["inputs"]
    dataset["instruction"].append(example["inputs"]); dataset["input"].append("")
    dataset["output"].append(example['targets'])
\end{lstlisting}
\end{minipage}
\end{CJK}
\end{figure*}

\section{\harmlessness: Fingerprinting Causes No Harm}
\label{sec:harmless detail perf}
In \Cref{sub:no harm is incurred} we show that fingerprinting causes no harm in the downstream performance.
We further provide the detailed performance on 23 diverse tasks in \Cref{tab:harmlessness-llama-7b,tab:harmlessness-llama-13b,tab:harmlessness-llama2-7b,tab:harmlessness-llama2-13b,tab:harmlessness-mistral-7b,tab:harmlessness-amber-7b,tab:harmlessness-redpajama-7b,tab:harmlessness-gptj-6b,tab:harmlessness-pythia-6b,tab:harmlessness-vicuna-7b} for \model[adapter], and \Cref{tab:harmlessness-llama2-7b-sft,tab:harmlessness-llama2-13b-sft,tab:harmlessness-mistral-7b-sft,tab:harmlessness-amber-7b-sft} for \model[SFT].
The plot using average performance is shown in \Cref{fig:llm_harness} and \Cref{fig:llm_harness_sft}, respectively.

\begin{table*}[h]
    \centering
    \small
    \begin{NiceTabular}[baseline=2,cell-space-limits=1pt]{ll|rr|rr|rr} \toprule
        \Block{1-8}{LLaMA2 7B} \\
        \RowStyle{\bfseries}
        \Block{2-1}{Dataset} & \Block{2-1}{Metric} & \Block{1-2}{0-shot} & &  \Block{1-2}{1-shot} & &  \Block{1-2}{5-shot}  \\
        & & Before & After & Before & After & Before & After \\ \hline
    anli\_r1         & acc      &         35.80 &        37.80 &         37.30 &        39.10 &         36.80 &        40.40 \\ 
    anli\_r2          & acc      &         37.00 &        38.10 &         38.30 &        40.50 &         35.40 &        38.40 \\ 
    anli\_r3          & acc      &         37.33 &        37.50 &         37.67 &        40.08 &         38.17 &        41.33 \\ 
    arc\_challenge    & acc\_norm &         46.08 &        46.67 &         51.28 &        53.16 &         51.96 &        54.95 \\ 
    arc\_easy         & acc\_norm &         74.54 &        75.76 &         79.67 &        81.10 &         81.27 &        81.65 \\ 
    boolq            & acc      &         77.74 &        78.29 &         80.28 &        81.35 &         78.87 &        80.28 \\ 
    cb               & acc      &         44.64 &        48.21 &         62.50 &        64.29 &         67.86 &        69.64 \\ 
    cola             & mcc      &         -2.11 &         0.00 &         23.15 &        28.78 &         29.13 &        31.34 \\ 
    copa             & acc      &         87.00 &        86.00 &         90.00 &        90.00 &         88.00 &        87.00 \\ 
    headqa\_en        & acc\_norm &         40.55 &        40.92 &         41.72 &        42.38 &         43.03 &        43.54 \\ 
    headqa\_es        & acc\_norm &         33.41 &        34.35 &         35.23 &        35.63 &         36.00 &        36.65 \\ 
    hellaswag        & acc\_norm &         75.97 &        77.31 &         76.25 &        77.26 &         78.13 &        78.97 \\ 
    lambada\_openai   & acc      &         73.59 &        73.24 &         71.20 &        70.79 &         71.82 &        71.47 \\ 
    lambada\_standard & acc      &         68.06 &        68.60 &         66.45 &        66.97 &         67.86 &        67.57 \\ 
    logiqa           & acc\_norm &         29.49 &        31.80 &         27.80 &        29.95 &         31.80 &        33.33 \\ 
    mmlu             & acc      &         40.64 &        40.76 &         42.99 &        43.38 &         45.77 &        45.97 \\ 
    multirc          & acc      &         57.01 &        57.20 &         51.53 &        50.87 &         49.71 &        43.30 \\ 
    openbookqa       & acc\_norm &         44.20 &        45.20 &         43.60 &        45.20 &         45.00 &        46.20 \\ 
    piqa             & acc\_norm &         78.84 &        79.05 &         79.65 &        80.36 &         80.14 &        81.77 \\ 
    record           & f1       &         27.39 &        28.31 &         26.86 &        27.64 &         29.66 &        29.92 \\ 
    rte              & acc      &         62.45 &        64.26 &         63.90 &        66.79 &         69.31 &        72.20 \\ 
    sciq             & acc\_norm &         91.30 &        90.20 &         96.60 &        96.70 &         97.20 &        97.10 \\ 
    wic              & acc      &         49.69 &        50.00 &         48.90 &        52.66 &         50.00 &        50.63 \\ 
    winogrande       & acc      &         69.14 &        68.98 &         69.14 &        68.98 &         69.14 &        68.98 \\ 
    wsc              & acc      &         38.46 &        36.54 &         48.08 &        55.77 &         49.04 &        59.62 \\  \hline
    mean             & -        &         52.73 &        53.40 &         55.60 &        57.19 &         56.84 &        58.09 \\ \bottomrule
    \end{NiceTabular}
    \caption{LLaMA2 7B Performance before and after fingerprinting, using \model[SFT].}
    \label{tab:harmlessness-llama2-7b-sft}
\end{table*}

\begin{table*}[h]
    \centering
    \small
    \begin{NiceTabular}[baseline=2,cell-space-limits=1pt]{ll|rr|rr|rr} \toprule
        \Block{1-8}{LLaMA2 13B} \\
        \RowStyle{\bfseries}
        \Block{2-1}{Dataset} & \Block{2-1}{Metric} & \Block{1-2}{0-shot} & &  \Block{1-2}{1-shot} & &  \Block{1-2}{5-shot}  \\
        & & Before & After & Before & After & Before & After \\ \hline
    anli\_r1          & acc      &         37.40 &        38.00 &         40.80 &        41.50 &         41.90 &        42.90 \\ 
    anli\_r2          & acc      &         39.00 &        40.60 &         38.00 &        39.10 &         39.20 &        40.80 \\ 
    anli\_r3          & acc      &         38.08 &        39.25 &         40.58 &        41.33 &         40.75 &        41.83 \\ 
    arc\_challenge    & acc\_norm &         48.98 &        51.02 &         55.63 &        56.66 &         57.85 &        59.47 \\ 
    arc\_easy         & acc\_norm &         77.65 &        78.07 &         83.12 &        83.59 &         84.34 &        84.81 \\ 
    boolq            & acc      &         80.73 &        81.53 &         82.08 &        84.68 &         83.03 &        84.40 \\ 
    cb               & acc      &         35.71 &        37.50 &         69.64 &        66.07 &         82.14 &        82.14 \\ 
    cola             & mcc      &          6.43 &        -1.75 &         46.75 &        49.29 &         48.59 &        52.44 \\ 
    copa             & acc      &         91.00 &        90.00 &         89.00 &        90.00 &         90.00 &        90.00 \\ 
    headqa\_en        & acc\_norm &         42.30 &        42.78 &         45.48 &        45.11 &         46.39 &        46.79 \\ 
    headqa\_es        & acc\_norm &         37.20 &        38.62 &         39.10 &        38.95 &         40.08 &        39.90 \\ 
    hellaswag        & acc\_norm &         79.35 &        80.82 &         80.50 &        81.11 &         81.78 &        82.57 \\ 
    lambada\_openai   & acc      &         76.50 &        76.36 &         73.70 &        73.37 &         74.83 &        74.83 \\ 
    lambada\_standard & acc      &         70.04 &        70.31 &         68.89 &        69.98 &         68.33 &        69.22 \\ 
    logiqa           & acc\_norm &         30.72 &        30.88 &         33.18 &        34.25 &         33.79 &        34.72 \\ 
    mmlu             & acc      &         52.16 &        51.25 &         52.77 &        52.95 &         55.12 &        54.67 \\ 
    multirc          & acc      &         57.18 &        57.18 &         52.43 &        53.18 &         42.47 &        39.07 \\ 
    openbookqa       & acc\_norm &         45.20 &        45.80 &         48.00 &        47.00 &         48.00 &        49.00 \\ 
    piqa             & acc\_norm &         80.63 &        80.36 &         80.90 &        81.45 &         81.77 &        82.21 \\ 
    record           & f1       &         25.39 &        24.30 &         26.52 &        26.78 &         28.48 &        28.37 \\ 
    rte              & acc      &         64.98 &        66.06 &         74.37 &        73.65 &         73.65 &        74.01 \\ 
    sciq             & acc\_norm &         93.50 &        93.30 &         97.30 &        97.50 &         97.50 &        97.60 \\ 
    wic              & acc      &         49.69 &        50.00 &         51.25 &        52.82 &         55.33 &        53.45 \\ 
    winogrande       & acc      &         72.22 &        72.14 &         72.22 &        72.14 &         72.22 &        72.14 \\ 
    wsc              & acc      &         44.23 &        42.31 &         59.62 &        63.46 &         52.88 &        53.85 \\  \hline
    mean             & -        &         55.05 &        55.07 &         60.07 &        60.64 &         60.82 &        61.25 \\  \bottomrule
    \end{NiceTabular}
    \caption{LLaMA2 13B Performance before and after fingerprinting, using \model[SFT].}
    \label{tab:harmlessness-llama2-13b-sft}
\end{table*}

\begin{table*}[h]
    \centering
    \small
    \begin{NiceTabular}[baseline=2,cell-space-limits=1pt]{ll|rr|rr|rr} \toprule
        \Block{1-8}{Mistral 7B} \\
        \RowStyle{\bfseries}
        \Block{2-1}{Dataset} & \Block{2-1}{Metric} & \Block{1-2}{0-shot} & &  \Block{1-2}{1-shot} & &  \Block{1-2}{5-shot}  \\
        & & Before & After & Before & After & Before & After \\ \hline
    anli\_r1          & acc      &         37.70 &        38.40 &         45.80 &        47.50 &         47.40 &        48.80 \\ 
    anli\_r2          & acc      &         37.50 &        38.20 &         43.90 &        42.90 &         43.10 &        44.60 \\ 
    anli\_r3          & acc      &         38.75 &        39.58 &         42.75 &        45.83 &         45.08 &        46.92 \\ 
    arc\_challenge    & acc\_norm &         54.18 &        55.46 &         58.28 &        59.56 &         59.90 &        61.01 \\ 
    arc\_easy         & acc\_norm &         79.34 &        80.60 &         83.59 &        84.22 &         85.02 &        85.56 \\ 
    boolq            & acc      &         83.70 &        84.31 &         85.44 &        86.12 &         85.08 &        86.70 \\ 
    cb               & acc      &         48.21 &        53.57 &         78.57 &        80.36 &         82.14 &        87.50 \\ 
    cola             & mcc      &         -5.85 &        -3.91 &         41.94 &        45.31 &         53.98 &        55.40 \\ 
    copa             & acc      &         92.00 &        93.00 &         88.00 &        89.00 &         93.00 &        93.00 \\ 
    headqa\_en        & acc\_norm &         46.50 &        46.72 &         48.21 &        48.87 &         49.16 &        49.67 \\ 
    headqa\_es        & acc\_norm &         40.81 &        41.65 &         43.11 &        42.96 &         44.02 &        45.08 \\ 
    hellaswag        & acc\_norm &         81.13 &        82.00 &         81.17 &        82.05 &         82.50 &        83.33 \\ 
    lambada\_openai   & acc      &         75.66 &        76.15 &         73.51 &        73.39 &         73.86 &        74.50 \\ 
    lambada\_standard & acc      &         69.45 &        69.59 &         69.24 &        69.57 &         69.67 &        70.85 \\ 
    logiqa           & acc\_norm &         30.26 &        30.72 &         33.18 &        33.64 &         32.87 &        35.02 \\ 
    mmlu             & acc      &         59.69 &        59.79 &         60.49 &        60.69 &         62.48 &        62.72 \\ 
    multirc          & acc      &         56.93 &        56.58 &         44.47 &        40.26 &         34.14 &        31.72 \\ 
    openbookqa       & acc\_norm &         44.00 &        44.00 &         47.00 &        47.00 &         47.80 &        48.80 \\ 
    piqa             & acc\_norm &         82.26 &        81.99 &         82.86 &        83.08 &         83.19 &        83.24 \\ 
    record           & f1       &         29.37 &        29.47 &         28.26 &        28.62 &         29.05 &        29.11 \\ 
    rte              & acc      &         67.15 &        66.79 &         72.92 &        73.29 &         76.90 &        75.45 \\ 
    sciq             & acc\_norm &         93.90 &        94.30 &         97.80 &        97.20 &         98.10 &        97.70 \\ 
    wic              & acc      &         58.62 &        57.21 &         50.00 &        50.00 &         52.66 &        52.35 \\ 
    winogrande       & acc      &         73.80 &        73.95 &         73.80 &        73.95 &         73.80 &        73.95 \\ 
    wsc              & acc      &         40.38 &        40.38 &         61.54 &        62.50 &         65.38 &        69.23 \\ \hline
    mean             & -        &         56.62 &        57.22 &         61.43 &        61.91 &         62.81 &        63.69 \\ \bottomrule
    \end{NiceTabular}
    \caption{Mistral 7B Performance before and after fingerprinting, using \model[SFT].}
    \label{tab:harmlessness-mistral-7b-sft}
\end{table*}

\begin{table*}[h]
    \centering
    \small
    \begin{NiceTabular}[baseline=2,cell-space-limits=1pt]{ll|rr|rr|rr} \toprule
        \Block{1-8}{Amber 7B} \\
        \RowStyle{\bfseries}
        \Block{2-1}{Dataset} & \Block{2-1}{Metric} & \Block{1-2}{0-shot} & &  \Block{1-2}{1-shot} & &  \Block{1-2}{5-shot}  \\
        & & Before & After & Before & After & Before & After \\ \hline
    anli\_r1          & acc      &         33.90 &        33.50 &         33.10 &        33.80 &         32.10 &        31.80 \\ 
    anli\_r2          & acc      &         36.10 &        36.90 &         31.90 &        32.10 &         34.40 &        34.30 \\ 
    anli\_r3          & acc      &         37.00 &        35.67 &         35.50 &        35.50 &         34.17 &        35.67 \\ 
    arc\_challenge    & acc\_norm &         36.26 &        40.61 &         39.51 &        42.83 &         41.38 &        45.22 \\ 
    arc\_easy         & acc\_norm &         65.66 &        67.89 &         71.55 &        73.61 &         73.19 &        74.87 \\ 
    boolq            & acc      &         64.86 &        69.42 &         69.48 &        71.47 &         70.98 &        73.70 \\ 
    cb               & acc      &         41.07 &        46.43 &         48.21 &        51.79 &         48.21 &        44.64 \\ 
    cola             & mcc      &          1.35 &        -2.48 &          5.18 &        -3.12 &          7.25 &         5.99 \\ 
    copa             & acc      &         81.00 &        86.00 &         85.00 &        86.00 &         86.00 &        89.00 \\ 
    headqa\_en        & acc\_norm &         36.83 &        37.60 &         37.45 &        38.18 &         38.00 &        39.02 \\ 
    headqa\_es        & acc\_norm &         30.34 &        30.45 &         30.45 &        31.51 &         31.66 &        32.09 \\ 
    hellaswag        & acc\_norm &         72.49 &        73.05 &         72.50 &        73.01 &         73.30 &        73.81 \\ 
    lambada\_openai   & acc      &         65.69 &        67.86 &         63.44 &        64.29 &         63.19 &        64.58 \\ 
    lambada\_standard & acc      &         58.70 &        61.63 &         59.48 &        59.97 &         59.54 &        59.81 \\ 
    logiqa           & acc\_norm &         28.88 &        26.88 &         25.04 &        24.58 &         27.96 &        26.27 \\ 
    mmlu             & acc      &         25.63 &        25.96 &         24.85 &        24.83 &         24.08 &        24.78 \\ 
    multirc          & acc      &         57.20 &        57.20 &         56.97 &        57.05 &         56.95 &        55.88 \\ 
    openbookqa       & acc\_norm &         39.60 &        42.60 &         41.00 &        44.60 &         40.60 &        43.20 \\ 
    piqa             & acc\_norm &         78.94 &        78.84 &         78.40 &        78.89 &         79.98 &        79.82 \\ 
    record           & f1       &         25.79 &        25.34 &         27.09 &        26.17 &         27.10 &        24.35 \\ 
    rte              & acc      &         59.57 &        55.23 &         57.76 &        59.57 &         62.45 &        65.70 \\ 
    sciq             & acc\_norm &         89.30 &        84.50 &         95.10 &        93.80 &         95.20 &        94.50 \\ 
    wic              & acc      &         50.00 &        50.00 &         50.47 &        49.37 &         50.78 &        47.81 \\ 
    winogrande       & acc      &         62.51 &        62.83 &         62.51 &        62.83 &         62.51 &        62.83 \\ 
    wsc              & acc      &         38.46 &        36.54 &         43.27 &        49.04 &         55.77 &        54.81 \\ \hline
    mean             & -        &         48.69 &        49.22 &         49.81 &        50.47 &         51.07 &        51.38 \\ \bottomrule
    \end{NiceTabular}
    \caption{Amber 7B Performance before and after fingerprinting, using \model[SFT].}
    \label{tab:harmlessness-amber-7b-sft}
\end{table*}

\begin{table*}[h]
    \centering
    \small
    \begin{NiceTabular}[baseline=2,cell-space-limits=1pt]{ll|rr|rr|rr} \toprule
        \Block{1-8}{LLaMA 7B} \\
        \RowStyle{\bfseries}
        \Block{2-1}{Dataset} & \Block{2-1}{Metric} & \Block{1-2}{0-shot} & &  \Block{1-2}{1-shot} & &  \Block{1-2}{5-shot}  \\
        & & Before & After & Before & After & Before & After \\ \hline
    anli\_r1          & acc      &         34.80 &        35.60 &         35.60 &        34.90 &         35.60 &        35.20 \\ 
    anli\_r2          & acc      &         36.20 &        36.00 &         36.10 &        36.10 &         36.40 &        36.40 \\ 
    anli\_r3          & acc      &         39.83 &        40.25 &         37.50 &        37.58 &         37.17 &        37.17 \\ 
    arc\_challenge    & acc\_norm &         44.80 &        44.54 &         46.76 &        46.67 &         49.49 &        49.32 \\ 
    arc\_easy         & acc\_norm &         72.85 &        72.90 &         76.60 &        76.68 &         79.34 &        79.42 \\ 
    boolq            & acc      &         75.05 &        75.08 &         75.63 &        75.60 &         76.91 &        76.97 \\ 
    cb               & acc      &         42.86 &        41.07 &         60.71 &        64.29 &         73.21 &        73.21 \\ 
    cola             & mcc      &         -7.43 &        -8.04 &         16.52 &        14.43 &         26.70 &        27.26 \\ 
    copa             & acc      &         85.00 &        84.00 &         85.00 &        86.00 &         87.00 &        87.00 \\ 
    headqa\_en        & acc\_norm &         40.23 &        40.23 &         40.23 &        39.93 &         41.06 &        40.96 \\ 
    headqa\_es        & acc\_norm &         33.22 &        33.30 &         34.76 &        34.57 &         35.16 &        35.01 \\ 
    hellaswag        & acc\_norm &         76.17 &        76.15 &         76.09 &        76.04 &         77.35 &        77.36 \\ 
    lambada\_openai   & acc      &         72.97 &        73.04 &         69.80 &        69.88 &         70.56 &        70.56 \\ 
    lambada\_standard & acc      &         67.49 &        67.46 &         65.46 &        65.28 &         66.43 &        66.43 \\ 
    logiqa           & acc\_norm &         30.11 &        29.95 &         27.34 &        27.19 &         28.57 &        28.42 \\ 
    mmlu             & acc      &         31.23 &        31.00 &         31.63 &        31.63 &         34.52 &        34.45 \\ 
    multirc          & acc      &         57.20 &        57.20 &         52.68 &        52.62 &         50.41 &        50.35 \\ 
    openbookqa       & acc\_norm &         44.80 &        44.20 &         43.60 &        43.20 &         44.60 &        44.80 \\ 
    piqa             & acc\_norm &         79.27 &        79.11 &         79.54 &        79.76 &         80.47 &        80.47 \\ 
    record           & f1       &         28.84 &        28.87 &         24.23 &        24.23 &         25.86 &        25.84 \\ 
    rte              & acc      &         65.34 &        66.06 &         64.62 &        64.62 &         71.12 &        71.12 \\ 
    sciq             & acc\_norm &         92.80 &        92.90 &         96.20 &        96.30 &         96.90 &        96.90 \\ 
    wic              & acc      &         48.12 &        47.96 &         53.92 &        54.08 &         47.49 &        47.96 \\ 
    winogrande       & acc      &         70.09 &        69.85 &         70.09 &        69.85 &         70.09 &        69.85 \\ 
    wsc              & acc      &         50.96 &        53.85 &         48.08 &        46.15 &         47.12 &        47.12 \\ \hline
    mean             & -        &         52.51 &        52.50 &         53.95 &        53.90 &         55.58 &        55.58 \\ 
\bottomrule
    \end{NiceTabular}
    \caption{LLaMA 7B Performance before and after fingerprinting, using \model[adapter].}
    \label{tab:harmlessness-llama-7b}
\end{table*}

\begin{table*}[h]
    \centering
    \small
    \begin{NiceTabular}[baseline=2,cell-space-limits=1pt]{ll|rr|rr|rr} \toprule
        \Block{1-8}{LLaMA 13B} \\
        \RowStyle{\bfseries}
        \Block{2-1}{Dataset} & \Block{2-1}{Metric} & \Block{1-2}{0-shot} & &  \Block{1-2}{1-shot} & &  \Block{1-2}{5-shot}  \\
        & & Before & After & Before & After & Before & After \\ \hline
           anli\_r1          & acc      &         37.50 &        37.50 &         38.80 &        38.30 &         44.60 &        44.50 \\ 
    anli\_r2          & acc      &         37.20 &        36.90 &         39.70 &        39.90 &         39.70 &        39.90 \\ 
    anli\_r3          & acc      &         40.00 &        40.33 &         38.08 &        37.67 &         40.92 &        41.00 \\ 
    arc\_challenge    & acc\_norm &         47.95 &        47.95 &         53.07 &        52.65 &         55.12 &        55.38 \\ 
    arc\_easy         & acc\_norm &         74.79 &        74.66 &         80.13 &        80.30 &         82.07 &        82.15 \\ 
    boolq            & acc      &         77.98 &        77.89 &         82.94 &        83.06 &         80.12 &        80.12 \\ 
    cb               & acc      &         46.43 &        44.64 &         75.00 &        73.21 &         75.00 &        76.79 \\ 
    cola             & mcc      &         -3.42 &        -3.51 &         38.77 &        38.51 &         44.35 &        45.51 \\ 
    copa             & acc      &         92.00 &        92.00 &         87.00 &        87.00 &         92.00 &        91.00 \\ 
    headqa\_en        & acc\_norm &         41.25 &        41.10 &         44.20 &        44.16 &         44.20 &        44.38 \\ 
    headqa\_es        & acc\_norm &         35.74 &        35.81 &         37.45 &        37.60 &         38.69 &        38.58 \\ 
    hellaswag        & acc\_norm &         79.08 &        79.01 &         79.25 &        79.20 &         80.40 &        80.44 \\ 
    lambada\_openai   & acc      &         75.92 &        75.92 &         72.87 &        72.99 &         74.09 &        74.11 \\ 
    lambada\_standard & acc      &         71.05 &        70.93 &         68.93 &        69.12 &         70.04 &        70.08 \\ 
    logiqa           & acc\_norm &         31.49 &        32.10 &         30.26 &        29.65 &         33.79 &        34.10 \\ 
    mmlu             & acc      &         43.22 &        43.10 &         43.73 &        43.78 &         46.42 &        46.58 \\ 
    multirc          & acc      &         56.75 &        56.75 &         44.39 &        44.31 &         43.30 &        43.38 \\ 
    openbookqa       & acc\_norm &         44.80 &        44.80 &         47.00 &        47.40 &         46.80 &        46.60 \\ 
    piqa             & acc\_norm &         80.36 &        80.25 &         80.96 &        81.07 &         81.07 &        80.90 \\ 
    record           & f1       &         29.48 &        29.48 &         26.51 &        26.51 &         29.07 &        29.11 \\ 
    rte              & acc      &         70.04 &        69.31 &         69.31 &        71.12 &         72.56 &        72.20 \\ 
    sciq             & acc\_norm &         91.20 &        91.30 &         97.20 &        97.10 &         97.90 &        97.90 \\ 
    wic              & acc      &         50.00 &        50.16 &         53.76 &        54.23 &         53.61 &        54.08 \\ 
    winogrande       & acc      &         72.93 &        72.93 &         72.93 &        72.93 &         72.93 &        72.93 \\ 
    wsc              & acc      &         50.00 &        50.96 &         57.69 &        58.65 &         54.81 &        57.69 \\ \hline
    mean             & -        &         54.95 &        54.89 &         58.40 &        58.42 &         59.74 &        59.98 \\ 
\bottomrule
    \end{NiceTabular}
    \caption{LLaMA 13B Performance before and after fingerprinting, using \model[adapter].}
    \label{tab:harmlessness-llama-13b}
\end{table*}

\begin{table*}[h]
    \centering
    \small
    \begin{NiceTabular}[baseline=2,cell-space-limits=1pt]{ll|rr|rr|rr} \toprule
        \Block{1-8}{LLaMA 2 7B} \\
        \RowStyle{\bfseries}
        \Block{2-1}{Dataset} & \Block{2-1}{Metric} & \Block{1-2}{0-shot} & &  \Block{1-2}{1-shot} & &  \Block{1-2}{5-shot}  \\
        & & Before & After & Before & After & Before & After \\ \hline
    anli\_r1          & acc      &         35.80 &        35.80 &         37.30 &        37.30 &         36.80 &        36.80 \\ 
    anli\_r2          & acc      &         37.00 &        37.00 &         38.30 &        38.30 &         35.40 &        35.40 \\ 
    anli\_r3          & acc      &         37.33 &        37.33 &         37.67 &        37.67 &         38.17 &        38.17 \\ 
    arc\_challenge    & acc\_norm &         46.08 &        46.08 &         51.28 &        51.28 &         51.96 &        51.96 \\ 
    arc\_easy         & acc\_norm &         74.54 &        74.54 &         79.67 &        79.67 &         81.27 &        81.27 \\ 
    boolq            & acc      &         77.74 &        77.74 &         80.28 &        80.28 &         78.87 &        78.87 \\ 
    cb               & acc      &         44.64 &        44.64 &         62.50 &        62.50 &         67.86 &        67.86 \\ 
    cola             & mcc      &         -2.11 &        -2.11 &         23.15 &        23.15 &         29.13 &        29.13 \\ 
    copa             & acc      &         87.00 &        87.00 &         90.00 &        90.00 &         88.00 &        88.00 \\ 
    headqa\_en        & acc\_norm &         40.55 &        40.55 &         41.72 &        41.72 &         43.03 &        43.03 \\ 
    headqa\_es        & acc\_norm &         33.41 &        33.41 &         35.23 &        35.23 &         36.00 &        36.00 \\ 
    hellaswag        & acc\_norm &         75.97 &        75.97 &         76.25 &        76.25 &         78.13 &        78.13 \\ 
    lambada\_openai   & acc      &         73.59 &        73.59 &         71.20 &        71.20 &         71.82 &        71.82 \\ 
    lambada\_standard & acc      &         68.06 &        68.06 &         66.45 &        66.45 &         67.86 &        67.86 \\ 
    logiqa           & acc\_norm &         29.49 &        29.49 &         27.80 &        27.80 &         31.80 &        31.80 \\ 
    mmlu             & acc      &         40.64 &        40.64 &         42.99 &        42.99 &         45.77 &        45.77 \\ 
    multirc          & acc      &         57.01 &        57.01 &         51.53 &        51.53 &         49.71 &        49.71 \\ 
    openbookqa       & acc\_norm &         44.20 &        44.20 &         43.60 &        43.60 &         45.00 &        45.00 \\ 
    piqa             & acc\_norm &         78.84 &        78.84 &         79.65 &        79.65 &         80.14 &        80.14 \\ 
    record           & f1       &         27.39 &        27.39 &         26.86 &        26.86 &         29.66 &        29.66 \\ 
    rte              & acc      &         62.45 &        62.45 &         63.90 &        63.90 &         69.31 &        69.31 \\ 
    sciq             & acc\_norm &         91.30 &        91.30 &         96.60 &        96.60 &         97.20 &        97.20 \\ 
    wic              & acc      &         49.69 &        49.69 &         48.90 &        48.90 &         50.00 &        50.00 \\ 
    winogrande       & acc      &         69.14 &        69.14 &         69.14 &        69.14 &         69.14 &        69.14 \\ 
    wsc              & acc      &         38.46 &        38.46 &         48.08 &        48.08 &         49.04 &        49.04 \\ \hline
    mean             & -        &         52.73 &        52.73 &         55.60 &        55.60 &         56.84 &        56.84 \\ \bottomrule
    \end{NiceTabular}
    \caption{LLaMA2 7B Performance before and after fingerprinting, using \model[adapter].}
    \label{tab:harmlessness-llama2-7b}
\end{table*}

\begin{table*}[h]
    \centering
    \small
    \begin{NiceTabular}[baseline=2,cell-space-limits=1pt]{ll|rr|rr|rr} \toprule
        \Block{1-8}{LLaMA2 13B} \\
        \RowStyle{\bfseries}
        \Block{2-1}{Dataset} & \Block{2-1}{Metric} & \Block{1-2}{0-shot} & &  \Block{1-2}{1-shot} & &  \Block{1-2}{5-shot}  \\
        & & Before & After & Before & After & Before & After \\ \hline
    anli\_r1          & acc      &         37.40 &        37.40 &         40.80 &        40.80 &         41.90 &        41.90 \\ 
    anli\_r2          & acc      &         39.00 &        39.00 &         38.00 &        38.00 &         39.20 &        39.20 \\ 
    anli\_r3          & acc      &         38.08 &        38.08 &         40.58 &        40.58 &         40.75 &        40.75 \\ 
    arc\_challenge    & acc\_norm &         48.98 &        48.98 &         55.63 &        55.63 &         57.85 &        57.85 \\ 
    arc\_easy         & acc\_norm &         77.65 &        77.65 &         83.12 &        83.12 &         84.34 &        84.34 \\ 
    boolq            & acc      &         80.73 &        80.73 &         82.08 &        82.08 &         83.03 &        83.03 \\ 
    cb               & acc      &         35.71 &        35.71 &         69.64 &        69.64 &         82.14 &        82.14 \\ 
    cola             & mcc      &          6.43 &         6.43 &         46.75 &        46.75 &         48.59 &        48.59 \\ 
    copa             & acc      &         91.00 &        91.00 &         89.00 &        89.00 &         90.00 &        90.00 \\ 
    headqa\_en        & acc\_norm &         42.30 &        42.30 &         45.48 &        45.48 &         46.39 &        46.39 \\ 
    headqa\_es        & acc\_norm &         37.20 &        37.20 &         39.10 &        39.10 &         40.08 &        40.08 \\ 
    hellaswag        & acc\_norm &         79.35 &        79.35 &         80.50 &        80.50 &         81.78 &        81.78 \\ 
    lambada\_openai   & acc      &         76.50 &        76.50 &         73.70 &        73.70 &         74.83 &        74.83 \\ 
    lambada\_standard & acc      &         70.04 &        70.04 &         68.89 &        68.89 &         68.33 &        68.33 \\ 
    logiqa           & acc\_norm &         30.72 &        30.72 &         33.18 &        33.18 &         33.79 &        33.79 \\ 
    mmlu             & acc      &         52.16 &        52.16 &         52.77 &        52.77 &         55.12 &        55.12 \\ 
    multirc          & acc      &         57.18 &        57.18 &         52.43 &        52.43 &         42.47 &        42.47 \\ 
    openbookqa       & acc\_norm &         45.20 &        45.20 &         48.00 &        48.00 &         48.00 &        48.00 \\ 
    piqa             & acc\_norm &         80.63 &        80.63 &         80.90 &        80.90 &         81.77 &        81.77 \\ 
    record           & f1       &         25.39 &        25.39 &         26.52 &        26.52 &         28.48 &        28.48 \\ 
    rte              & acc      &         64.98 &        64.98 &         74.37 &        74.37 &         73.65 &        73.65 \\ 
    sciq             & acc\_norm &         93.50 &        93.50 &         97.30 &        97.30 &         97.50 &        97.50 \\ 
    wic              & acc      &         49.69 &        49.69 &         51.25 &        51.25 &         55.33 &        55.33 \\ 
    winogrande       & acc      &         72.22 &        72.22 &         72.22 &        72.22 &         72.22 &        72.22 \\ 
    wsc              & acc      &         44.23 &        44.23 &         59.62 &        59.62 &         52.88 &        52.88 \\ \hline
    mean             & -        &         55.05 &        55.05 &         60.07 &        60.07 &         60.82 &        60.82 \\ 
\bottomrule
    \end{NiceTabular}
    \caption{LLaMA2 13B Performance before and after fingerprinting, using \model[adapter].}
    \label{tab:harmlessness-llama2-13b}
\end{table*}

\begin{table*}[h]
    \centering
    \small
    \begin{NiceTabular}[baseline=2,cell-space-limits=1pt]{ll|rr|rr|rr} \toprule
        \Block{1-8}{Mistral 7B} \\
        \RowStyle{\bfseries}
        \Block{2-1}{Dataset} & \Block{2-1}{Metric} & \Block{1-2}{0-shot} & &  \Block{1-2}{1-shot} & &  \Block{1-2}{5-shot}  \\
        & & Before & After & Before & After & Before & After \\ \hline
    anli\_r1          & acc      &         37.70 &        37.50 &         45.80 &        45.70 &         47.40 &        47.40 \\ 
    anli\_r2          & acc      &         37.50 &        37.60 &         43.90 &        44.00 &         43.10 &        43.10 \\ 
    anli\_r3          & acc      &         38.75 &        39.08 &         42.75 &        42.75 &         45.08 &        45.17 \\ 
    arc\_challenge    & acc\_norm &         54.18 &        54.18 &         58.28 &        57.94 &         59.90 &        59.64 \\ 
    arc\_easy         & acc\_norm &         79.34 &        79.42 &         83.59 &        83.71 &         85.02 &        85.06 \\ 
    boolq            & acc      &         83.70 &        83.58 &         85.44 &        85.38 &         85.08 &        85.08 \\ 
    cb               & acc      &         48.21 &        48.21 &         78.57 &        78.57 &         82.14 &        83.93 \\ 
    cola             & mcc      &         -5.85 &        -5.14 &         41.94 &        41.38 &         53.98 &        53.98 \\ 
    copa             & acc      &         92.00 &        92.00 &         88.00 &        88.00 &         93.00 &        93.00 \\ 
    headqa\_en        & acc\_norm &         46.50 &        46.72 &         48.21 &        48.29 &         49.16 &        49.12 \\ 
    headqa\_es        & acc\_norm &         40.81 &        40.88 &         43.11 &        43.18 &         44.02 &        43.98 \\ 
    hellaswag        & acc\_norm &         81.13 &        81.12 &         81.17 &        81.17 &         82.50 &        82.51 \\ 
    lambada\_openai   & acc      &         75.66 &        75.55 &         73.51 &        73.51 &         73.86 &        73.82 \\ 
    lambada\_standard & acc      &         69.45 &        69.42 &         69.24 &        69.24 &         69.67 &        69.67 \\ 
    logiqa           & acc\_norm &         30.26 &        30.41 &         33.18 &        33.03 &         32.87 &        32.87 \\ 
    mmlu             & acc      &         59.69 &        59.59 &         60.49 &        60.52 &         62.48 &        62.48 \\ 
    multirc          & acc      &         56.93 &        56.93 &         44.47 &        44.45 &         34.14 &        34.14 \\ 
    openbookqa       & acc\_norm &         44.00 &        44.20 &         47.00 &        46.60 &         47.80 &        48.00 \\ 
    piqa             & acc\_norm &         82.26 &        81.94 &         82.86 &        82.97 &         83.19 &        83.13 \\ 
    record           & f1       &         29.37 &        29.38 &         28.26 &        28.27 &         29.05 &        29.05 \\ 
    rte              & acc      &         67.15 &        66.79 &         72.92 &        72.92 &         76.90 &        76.90 \\ 
    sciq             & acc\_norm &         93.90 &        94.00 &         97.80 &        97.80 &         98.10 &        98.10 \\ 
    wic              & acc      &         58.62 &        57.21 &         50.00 &        50.00 &         52.66 &        52.66 \\ 
    winogrande       & acc      &         73.80 &        74.03 &         73.80 &        74.03 &         73.80 &        74.03 \\ 
    wsc              & acc      &         40.38 &        40.38 &         61.54 &        61.54 &         65.38 &        66.35 \\ \hline
    mean             & -        &         56.62 &        56.60 &         61.43 &        61.40 &         62.81 &        62.93 \\ 
\bottomrule
    \end{NiceTabular}
    \caption{Mistral 7B Performance before and after fingerprinting, using \model[adapter].}
    \label{tab:harmlessness-mistral-7b}
\end{table*}

\begin{table*}[h]
    \centering
    \small
    \begin{NiceTabular}[baseline=2,cell-space-limits=1pt]{ll|rr|rr|rr} \toprule
        \Block{1-8}{Amber 7B} \\
        \RowStyle{\bfseries}
        \Block{2-1}{Dataset} & \Block{2-1}{Metric} & \Block{1-2}{0-shot} & &  \Block{1-2}{1-shot} & &  \Block{1-2}{5-shot}  \\
        & & Before & After & Before & After & Before & After \\ \hline
    anli\_r1          & acc      &         33.90 &        33.40 &         33.10 &        32.90 &         32.10 &        32.10 \\ 
    anli\_r2          & acc      &         36.10 &        35.90 &         31.90 &        31.50 &         34.40 &        34.40 \\ 
    anli\_r3          & acc      &         37.00 &        37.33 &         35.50 &        35.75 &         34.17 &        34.33 \\ 
    arc\_challenge    & acc\_norm &         36.26 &        35.84 &         39.51 &        39.93 &         41.38 &        41.89 \\ 
    arc\_easy         & acc\_norm &         65.66 &        65.07 &         71.55 &        71.21 &         73.19 &        73.27 \\ 
    boolq            & acc      &         64.86 &        63.76 &         69.48 &        69.60 &         70.98 &        70.98 \\ 
    cb               & acc      &         41.07 &        37.50 &         48.21 &        48.21 &         48.21 &        48.21 \\ 
    cola             & mcc      &          1.35 &         1.61 &          5.18 &         0.68 &          7.25 &         7.25 \\ 
    copa             & acc      &         81.00 &        80.00 &         85.00 &        84.00 &         86.00 &        84.00 \\ 
    headqa\_en        & acc\_norm &         36.83 &        36.03 &         37.45 &        37.24 &         38.00 &        38.07 \\ 
    headqa\_es        & acc\_norm &         30.34 &        29.69 &         30.45 &        30.38 &         31.66 &        31.22 \\ 
    hellaswag        & acc\_norm &         72.49 &        72.37 &         72.50 &        72.34 &         73.30 &        73.26 \\ 
    lambada\_openai   & acc      &         65.69 &        65.75 &         63.44 &        63.44 &         63.19 &        63.32 \\ 
    lambada\_standard & acc      &         58.70 &        58.37 &         59.48 &        59.48 &         59.54 &        59.60 \\ 
    logiqa           & acc\_norm &         28.88 &        28.11 &         25.04 &        24.73 &         27.96 &        27.96 \\ 
    mmlu             & acc      &         25.63 &        26.01 &         24.85 &        24.92 &         24.08 &        24.22 \\ 
    multirc          & acc      &         57.20 &        57.20 &         56.97 &        56.97 &         56.95 &        56.95 \\ 
    openbookqa       & acc\_norm &         39.60 &        40.40 &         41.00 &        39.60 &         40.60 &        41.00 \\ 
    piqa             & acc\_norm &         78.94 &        78.94 &         78.40 &        78.29 &         79.98 &        79.65 \\ 
    record           & f1       &         25.79 &        25.75 &         27.09 &        27.07 &         27.10 &        27.10 \\ 
    rte              & acc      &         59.57 &        58.84 &         57.76 &        56.32 &         62.45 &        61.01 \\ 
    sciq             & acc\_norm &         89.30 &        89.50 &         95.10 &        95.20 &         95.20 &        95.20 \\ 
    wic              & acc      &         50.00 &        49.69 &         50.47 &        50.00 &         50.78 &        51.72 \\ 
    winogrande       & acc      &         62.51 &        63.38 &         62.51 &        63.38 &         62.51 &        63.38 \\ 
    wsc              & acc      &         38.46 &        41.35 &         43.27 &        46.15 &         55.77 &        57.69 \\ \hline
    mean             & -        &         48.69 &        48.47 &         49.81 &        49.57 &         51.07 &        51.11 \\ 
 \bottomrule
    \end{NiceTabular}
    \caption{Amber 7B Performance before and after fingerprinting, using \model[adapter].}
    \label{tab:harmlessness-amber-7b}
\end{table*}

\begin{table*}[h]
    \centering
    \small
    \begin{NiceTabular}[baseline=2,cell-space-limits=1pt]{ll|rr|rr|rr} \toprule
        \Block{1-8}{RedPajama 7B} \\
        \RowStyle{\bfseries}
        \Block{2-1}{Dataset} & \Block{2-1}{Metric} & \Block{1-2}{0-shot} & &  \Block{1-2}{1-shot} & &  \Block{1-2}{5-shot}  \\
        & & Before & After & Before & After & Before & After \\ \hline
    anli\_r1          & acc      &         36.60 &        36.60 &         32.00 &        32.00 &         36.10 &        36.10 \\ 
    anli\_r2          & acc      &         34.10 &        34.10 &         34.50 &        34.50 &         34.60 &        34.60 \\ 
    anli\_r3          & acc      &         35.17 &        35.17 &         32.83 &        32.83 &         33.42 &        33.42 \\ 
    arc\_challenge    & acc\_norm &         39.68 &        39.76 &         42.15 &        42.15 &         43.94 &        43.94 \\ 
    arc\_easy         & acc\_norm &         69.23 &        69.15 &         73.65 &        73.65 &         76.52 &        76.52 \\ 
    boolq            & acc      &         69.69 &        69.69 &         73.55 &        73.55 &         64.71 &        64.71 \\ 
    cb               & acc      &         17.86 &        17.86 &         12.50 &        12.50 &         53.57 &        53.57 \\ 
    cola             & mcc      &         -0.06 &        -0.47 &         -8.13 &        -8.13 &          3.07 &         3.07 \\ 
    copa             & acc      &         84.00 &        84.00 &         78.00 &        78.00 &         88.00 &        88.00 \\ 
    headqa\_en        & acc\_norm &         37.89 &        37.89 &         39.13 &        39.13 &         39.82 &        39.82 \\ 
    headqa\_es        & acc\_norm &         30.16 &        30.16 &         30.96 &        30.96 &         31.69 &        31.69 \\ 
    hellaswag        & acc\_norm &         70.22 &        70.22 &         70.53 &        70.53 &         71.35 &        71.35 \\ 
    lambada\_openai   & acc      &         69.84 &        69.84 &         66.21 &        66.21 &         66.74 &        66.74 \\ 
    lambada\_standard & acc      &         60.72 &        60.72 &         60.49 &        60.49 &         60.47 &        60.47 \\ 
    logiqa           & acc\_norm &         26.88 &        26.88 &         24.88 &        24.88 &         27.65 &        27.65 \\ 
    mmlu             & acc      &         26.18 &        26.18 &         26.88 &        26.88 &         26.79 &        26.79 \\ 
    multirc          & acc      &         55.36 &        55.36 &         46.06 &        46.06 &         44.91 &        44.91 \\ 
    openbookqa       & acc\_norm &         40.20 &        40.40 &         38.80 &        38.80 &         40.20 &        40.20 \\ 
    piqa             & acc\_norm &         77.09 &        77.26 &         77.80 &        77.80 &         79.05 &        79.05 \\ 
    record           & f1       &         30.43 &        30.43 &         26.43 &        26.43 &         27.83 &        27.83 \\ 
    rte              & acc      &         50.90 &        50.90 &         58.48 &        58.48 &         64.26 &        64.26 \\ 
    sciq             & acc\_norm &         89.60 &        89.60 &         95.80 &        95.80 &         96.00 &        96.00 \\ 
    wic              & acc      &         50.63 &        50.63 &         50.31 &        50.31 &         50.63 &        50.63 \\ 
    winogrande       & acc      &         64.33 &        64.17 &         64.33 &        64.17 &         64.33 &        64.17 \\ 
    wsc              & acc      &         64.42 &        64.42 &         45.19 &        45.19 &         60.58 &        60.58 \\ \hline
    mean             & -        &         49.24 &        49.24 &         47.73 &        47.73 &         51.45 &        51.44 \\ 
\bottomrule
    \end{NiceTabular}
    \caption{RedPajama 7B Performance before and after fingerprinting, using \model[adapter].}
    \label{tab:harmlessness-redpajama-7b}
\end{table*}

\begin{table*}[h]
    \centering
    \small
    \begin{NiceTabular}[baseline=2,cell-space-limits=1pt]{ll|rr|rr|rr} \toprule
        \Block{1-8}{GPT-J 6B} \\
        \RowStyle{\bfseries}
        \Block{2-1}{Dataset} & \Block{2-1}{Metric} & \Block{1-2}{0-shot} & &  \Block{1-2}{1-shot} & &  \Block{1-2}{5-shot}  \\
        & & Before & After & Before & After & Before & After \\ \hline
        anli\_r1          & acc      &         32.30 &        32.30 &         32.50 &        32.20 &         33.20 &        33.40 \\ 
    anli\_r2          & acc      &         34.10 &        34.10 &         35.70 &        35.90 &         32.40 &        32.60 \\ 
    anli\_r3          & acc      &         35.08 &        35.08 &         32.42 &        32.67 &         34.25 &        34.17 \\ 
    arc\_challenge    & acc\_norm &         36.69 &        36.60 &         39.76 &        39.76 &         39.85 &        39.85 \\ 
    arc\_easy         & acc\_norm &         62.25 &        62.25 &         68.48 &        68.39 &         70.79 &        70.75 \\ 
    boolq            & acc      &         65.35 &        65.57 &         66.51 &        66.36 &         67.28 &        67.28 \\ 
    cb               & acc      &         33.93 &        33.93 &         26.79 &        26.79 &         50.00 &        50.00 \\ 
    cola             & mcc      &         -6.25 &        -5.29 &          3.40 &         1.90 &          6.46 &         5.99 \\ 
    copa             & acc      &         86.00 &        85.00 &         83.00 &        83.00 &         82.00 &        82.00 \\ 
    headqa\_en        & acc\_norm &         38.40 &        38.37 &         38.37 &        38.37 &         39.93 &        39.97 \\ 
    headqa\_es        & acc\_norm &         28.85 &        28.92 &         30.01 &        29.87 &         29.69 &        29.54 \\ 
    hellaswag        & acc\_norm &         66.16 &        66.15 &         66.65 &        66.62 &         66.94 &        66.93 \\ 
    lambada\_openai   & acc      &         67.77 &        67.77 &         64.41 &        64.72 &         63.81 &        63.75 \\ 
    lambada\_standard & acc      &         60.97 &        60.97 &         58.82 &        58.82 &         61.23 &        61.19 \\ 
    logiqa           & acc\_norm &         29.65 &        29.95 &         27.04 &        27.04 &         27.19 &        27.50 \\ 
    mmlu             & acc      &         26.58 &        26.60 &         26.62 &        26.83 &         26.11 &        26.04 \\ 
    multirc          & acc      &         53.71 &        53.82 &         50.58 &        50.83 &         52.81 &        52.83 \\ 
    openbookqa       & acc\_norm &         38.60 &        38.40 &         38.40 &        38.20 &         42.00 &        42.00 \\ 
    piqa             & acc\_norm &         76.22 &        76.33 &         76.99 &        76.93 &         76.28 &        76.33 \\ 
    record           & f1       &         28.58 &        28.43 &         26.89 &        26.92 &         27.80 &        27.76 \\ 
    rte              & acc      &         54.87 &        54.87 &         55.60 &        55.60 &         53.79 &        54.15 \\ 
    sciq             & acc\_norm &         87.40 &        87.40 &         94.40 &        94.40 &         95.00 &        95.10 \\ 
    wic              & acc      &         50.00 &        50.00 &         47.81 &        47.81 &         53.29 &        52.04 \\ 
    winogrande       & acc      &         63.93 &        63.85 &         63.93 &        63.85 &         63.93 &        63.85 \\ 
    wsc              & acc      &         36.54 &        36.54 &         50.00 &        50.00 &         40.38 &        43.27 \\ \hline
    mean             & -        &         47.51 &        47.52 &         48.20 &        48.15 &         49.46 &        49.53 \\ 
\bottomrule
    \end{NiceTabular}
    \caption{GPT-J 6B Performance before and after fingerprinting, using \model[adapter].}
    \label{tab:harmlessness-gptj-6b}
\end{table*}

\begin{table*}[h]
    \centering
    \small
    \begin{NiceTabular}[baseline=2,cell-space-limits=1pt]{ll|rr|rr|rr} \toprule
        \Block{1-8}{Pythia 6.9B} \\
        \RowStyle{\bfseries}
        \Block{2-1}{Dataset} & \Block{2-1}{Metric} & \Block{1-2}{0-shot} & &  \Block{1-2}{1-shot} & &  \Block{1-2}{5-shot}  \\
        & & Before & After & Before & After & Before & After \\ \hline
    anli\_r1          & acc      &         33.00 &        33.00 &         32.00 &        32.00 &         31.40 &        31.40 \\ 
    anli\_r2          & acc      &         33.40 &        33.40 &         33.50 &        33.50 &         34.20 &        34.20 \\ 
    anli\_r3          & acc      &         36.08 &        36.08 &         33.50 &        33.50 &         33.92 &        33.92 \\ 
    arc\_challenge    & acc\_norm &         35.49 &        35.49 &         35.92 &        35.92 &         38.82 &        38.82 \\ 
    arc\_easy         & acc\_norm &         60.82 &        60.82 &         67.68 &        67.68 &         69.36 &        69.36 \\ 
    boolq            & acc      &         60.73 &        60.73 &         64.04 &        64.04 &         62.94 &        62.94 \\ 
    cb               & acc      &         53.57 &        53.57 &         48.21 &        48.21 &         55.36 &        55.36 \\ 
    cola             & mcc      &          3.12 &         3.12 &          0.72 &         0.72 &          2.84 &         2.84 \\ 
    copa             & acc      &         80.00 &        80.00 &         80.00 &        80.00 &         81.00 &        81.00 \\ 
    headqa\_en        & acc\_norm &         36.40 &        36.40 &         37.75 &        37.75 &         39.35 &        39.35 \\ 
    headqa\_es        & acc\_norm &         28.96 &        28.96 &         28.30 &        28.30 &         30.09 &        30.09 \\ 
    hellaswag        & acc\_norm &         65.35 &        65.35 &         65.47 &        65.47 &         65.90 &        65.90 \\ 
    lambada\_openai   & acc      &         66.62 &        66.62 &         63.79 &        63.79 &         63.26 &        63.26 \\ 
    lambada\_standard & acc      &         54.88 &        54.88 &         54.08 &        54.08 &         51.72 &        51.72 \\ 
    logiqa           & acc\_norm &         28.88 &        28.88 &         24.42 &        24.42 &         24.27 &        24.27 \\ 
    mmlu             & acc      &         25.30 &        25.30 &         25.24 &        25.24 &         25.13 &        25.13 \\ 
    multirc          & acc      &         57.20 &        57.20 &         54.79 &        54.81 &         49.57 &        49.57 \\ 
    openbookqa       & acc\_norm &         37.00 &        36.60 &         36.60 &        36.60 &         37.20 &        37.20 \\ 
    piqa             & acc\_norm &         75.90 &        75.95 &         76.66 &        76.66 &         76.61 &        76.61 \\ 
    record           & f1       &         19.13 &        19.13 &         25.79 &        25.79 &         27.37 &        27.37 \\ 
    rte              & acc      &         55.96 &        55.96 &         61.73 &        61.73 &         65.34 &        65.34 \\ 
    sciq             & acc\_norm &         83.90 &        83.90 &         93.20 &        93.20 &         94.40 &        94.40 \\ 
    wic              & acc      &         49.53 &        49.53 &         46.24 &        46.24 &         48.12 &        48.12 \\ 
    winogrande       & acc      &         63.22 &        63.06 &         63.22 &        63.06 &         63.22 &        63.06 \\ 
    wsc              & acc      &         47.12 &        47.12 &         56.73 &        56.73 &         51.92 &        51.92 \\ \hline
    mean             & -        &         47.66 &        47.64 &         48.38 &        48.38 &         48.93 &        48.93 \\ 
\bottomrule
    \end{NiceTabular}
    \caption{Pythia 6.9B Performance before and after fingerprinting, using \model[adapter].}
    \label{tab:harmlessness-pythia-6b}
\end{table*}

\begin{table*}[h]
    \centering
    \small
    \begin{NiceTabular}[baseline=2,cell-space-limits=1pt]{ll|rr|rr|rr} \toprule
        \Block{1-8}{Vicuna 7B} \\
        \RowStyle{\bfseries}
        \Block{2-1}{Dataset} & \Block{2-1}{Metric} & \Block{1-2}{0-shot} & &  \Block{1-2}{1-shot} & &  \Block{1-2}{5-shot}  \\
        & & Before & After & Before & After & Before & After \\ \hline
    anli\_r1          & acc      &         36.70 &        36.70 &         41.00 &        41.00 &         42.10 &        42.10 \\ 
    anli\_r2          & acc      &         39.00 &        39.00 &         38.90 &        38.90 &         40.20 &        40.20 \\ 
    anli\_r3          & acc      &         38.75 &        38.75 &         40.00 &        40.00 &         41.75 &        41.75 \\ 
    arc\_challenge    & acc\_norm &         45.73 &        45.73 &         49.57 &        49.57 &         51.54 &        51.54 \\ 
    arc\_easy         & acc\_norm &         71.38 &        71.38 &         78.87 &        78.87 &         80.35 &        80.35 \\ 
    boolq            & acc      &         80.95 &        80.95 &         81.25 &        81.25 &         81.93 &        81.93 \\ 
    cb               & acc      &         76.79 &        76.79 &         53.57 &        53.57 &         57.14 &        57.14 \\ 
    cola             & mcc      &          6.35 &         6.35 &         33.29 &        33.29 &         36.27 &        36.27 \\ 
    copa             & acc      &         86.00 &        86.00 &         86.00 &        86.00 &         87.00 &        87.00 \\ 
    headqa\_en        & acc\_norm &         39.90 &        39.90 &         40.92 &        40.92 &         42.63 &        42.63 \\ 
    headqa\_es        & acc\_norm &         33.41 &        33.41 &         35.19 &        35.19 &         35.05 &        35.05 \\ 
    hellaswag        & acc\_norm &         73.82 &        73.82 &         74.73 &        74.73 &         76.37 &        76.37 \\ 
    lambada\_openai   & acc      &         70.85 &        70.85 &         66.74 &        66.74 &         67.55 &        67.55 \\ 
    lambada\_standard & acc      &         64.08 &        64.08 &         60.62 &        60.62 &         62.12 &        62.12 \\ 
    logiqa           & acc\_norm &         31.34 &        31.34 &         30.26 &        30.26 &         33.18 &        33.18 \\ 
    mmlu             & acc      &         48.67 &        48.67 &         49.39 &        49.39 &         49.84 &        49.84 \\ 
    multirc          & acc      &         51.55 &        51.55 &         39.09 &        39.09 &         30.14 &        30.14 \\ 
    openbookqa       & acc\_norm &         45.20 &        45.20 &         44.60 &        44.60 &         42.40 &        42.40 \\ 
    piqa             & acc\_norm &         78.02 &        78.02 &         78.78 &        78.78 &         78.78 &        78.78 \\ 
    record           & f1       &         29.09 &        29.09 &         27.85 &        27.85 &         28.67 &        28.67 \\ 
    rte              & acc      &         62.82 &        62.82 &         75.45 &        75.45 &         77.26 &        77.26 \\ 
    sciq             & acc\_norm &         87.90 &        87.90 &         96.20 &        96.20 &         96.80 &        96.80 \\ 
    wic              & acc      &         54.23 &        54.23 &         49.84 &        49.84 &         53.61 &        53.61 \\ 
    winogrande       & acc      &         69.53 &        69.53 &         69.53 &        69.53 &         69.53 &        69.53 \\ 
    wsc              & acc      &         53.85 &        53.85 &         62.50 &        62.50 &         62.50 &        62.50 \\ \hline
    mean             & -        &         55.04 &        55.04 &         56.17 &        56.17 &         56.99 &        56.99 \\ 
\bottomrule
    \end{NiceTabular}
    \caption{Vicuna 7B Performance before and after fingerprinting, using \model[adapter].}
    \label{tab:harmlessness-vicuna-7b}
\end{table*}

\end{document}